\documentclass[twocolumn]{aastex63}

\received{2022 December 19}
\revised{2023 February 20}
\accepted{2023 March 10}

\submitjournal{The Astrophysical Journal}

\shorttitle{Are solar flares threshold-driven?}
\shortauthors{Carlin et al.}

\usepackage{xcolor} % for red text colouring
\usepackage{booktabs}	% Better tables
\usepackage{array}		% More tables options
\usepackage[inline, shortlabels]{enumitem}	% Options for lists
\usepackage{amsmath}

\newcommand{\etadx}{\eta\left[\Delta X_{i}\,|\,X(t_i^-)\right]}
\newcommand{\td}{\textrm{d}} % roman "d" for infinitesimals in integrals
\newcommand{\xc}{X_\textrm{cr}} % critical stress
\begin{document}
\title{A statistical search for a uniform trigger threshold in solar flares from individual active regions}

\correspondingauthor{Julian B. Carlin}
\email{jcarlin@student.unimelb.edu.au}

\author[0000-0001-5694-0809]{Julian B. Carlin}
\affiliation{School of Physics, University of Melbourne, Parkville, VIC 3010, Australia}
\affiliation{OzGrav, University of Melbourne, Parkville, VIC 3010, Australia}

\author{Andrew Melatos}
\affiliation{School of Physics, University of Melbourne, Parkville, VIC 3010, Australia}
\affiliation{OzGrav, University of Melbourne, Parkville, VIC 3010, Australia}

\author[0000-0001-5100-2354]{Michael S. Wheatland}
\affiliation{Sydney Institute for Astronomy, School of Physics, The University of Sydney, NSW 2006, Australia}

\begin{abstract}
Solar flares result from the sudden release of energy deposited by sub-photospheric motions into the magnetic field of the corona. The deposited energy accumulates secularly between events. One may interpret the observed event statistics as resulting from a state-dependent Poisson process, in which the instantaneous flare rate is a function of the stress in the system, and a flare becomes certain as the stress approaches a threshold set by the micro-physics of the flare trigger. If the system is driven fast, and if the threshold is static and uniform globally, a cross-correlation is predicted between the size of a flare and the forward waiting time to the next flare. This cross-correlation is broadly absent from the \emph{Geostationary Operational Environmental Satellite} (\emph{GOES}) soft X-ray flare database. One also predicts higher cross-correlations in active regions where the shapes of the waiting time and size distributions match. Again there is no evidence for such an association in the \emph{GOES} data. The data imply at least one of the following: \begin{enumerate*}[i)] \item the threshold at which a flare is triggered varies in time; \item the rate at which energy is driven into active regions varies in time; \item historical flare catalogs are incomplete; or \item the description of solar flares as resulting from a build-up and release of energy, once a threshold is reached, is incomplete. \end{enumerate*}
\end{abstract}


\section{Introduction} \label{sec:intro}
Broad consensus exists that the micro-physical process triggering individual solar flares is magnetic reconnection in the corona. A reconnection event becomes more likely to occur the more magnetic energy accumulates in the time between flares \citep{Priest2002, Benz2016}. This phenomenological class of stress-relax model was popularized by \citet{Rosner1978}, and expounded by \citet{Wheatland1998, Wheatland2000corr, Wheatland2008, Kanazir2010, Hudson2019}, among others. A complementary phenomenological description is the avalanche or self-organized criticality model \citep{Lu1991, Lu1993, Lu1995}, inspired by the canonical sandpiles of \citet{Bak1987, Bak1988}. These two classes of model are broadly compatible, but make different predictions about some long-term statistical observables \citep{Lu1995a, Boffetta1999, Wheatland2000corr, Lippiello2010, Farhang2019, Aschwanden2021a}. 

Upon aggregating historical datasets, the probability density function (PDF) of the energy released in solar flares is found to be a power law or log-normal over multiple decades \citep{Rosner1978, Lu1991, Verbeeck2019}. The PDF of time intervals between successive flares in the same active region, henceforth termed waiting times, is less clear-cut. \citet{Wheatland2000delt} found evidence that the power-law-like shape of the tail of the waiting time PDF found by \citet{Boffetta1999} is explained by a sum of exponentials, with individual rates themselves drawn from an exponential PDF. This interpretation is further developed by \citet{Aschwanden2021}. Flaring rates that vary in time are also noted by \citet{Lepreti2001} and \citet{Gorobets2012}. Cross-correlations between the size of a flare and the subsequent (preceding) waiting time, henceforth termed forward (backward) cross-correlations, are a key differentiator between the stress-relax and avalanche descriptions --- the latter predicts no size--waiting-time cross-correlations \citep{Jensen1998}. Forward and backward correlations are broadly absent from solar flares datasets \citep{Biesecker1994, Hudson1998, Crosby1998, Wheatland2000corr, Hudson2020}, with the exception of strong forward cross-correlations found in two active regions \citep{Hudson2019}.

Rotational glitches in rotation-powered pulsars \citep{Lyne2012, Haskell2015} are an analogous astrophysical phenomenon to solar flares, in the limited sense that they are consistent with a stress-relax process even though they do not involve magnetic reconnection as far as one knows \citep{Aschwanden2018}. While the exact process that triggers a glitch is unknown, most models are encompassed by the fundamental idea that ``stress'' (possibly differential rotation or elastic deformation) builds up secularly between glitches, and is released sporadically and partially at a glitch. The instantaneous glitch rate is assumed to grow with the stress in the system. This idea is formalized phenomenologically in the state-dependent Poisson (SDP) process popularized by \citet{Fulgenzi2017}. Precise, falsifiable predictions about size and waiting time PDFs, auto-, and cross-correlations, as well as comparisons to current datasets, show the power and flexibility of the SDP model in the neutron star context \citep{Melatos2018, Carlin2019quasi, Carlin2019ac, Melatos2019}. The same falsifiable predictions also deliver new physical insights when applied to solar flare data, as we show in this paper. 

Our goal in this paper is to search the \emph{Geostationary Operational Environmental Satellite} (\emph{GOES}) soft X-ray flare database for signatures of a threshold-driven stress-relax process. We do this by de-aggregating the data from different active regions, and studying summary statistics of flare waiting times and sizes. In Section~\ref{sec:sdp} we outline the SDP framework, and how it maps to solar flares. In Section~\ref{sec:pred} we explore various regimes of the SDP process. We also specify precise, falsifiable tests for the question of whether solar flares are triggered once the energy reaches a static threshold, if it obeys a stress-relax process. The \emph{GOES} soft X-ray dataset to which we apply these tests is described in Section~\ref{sec:data}. In Section~\ref{sec:disagg_general} we look for associations between matching waiting time and size PDFs and high size--waiting-time cross-correlations, as predicted for the SDP process. We conclude with a discussion of the microphysical implications of the data analysis in Section~\ref{sec:conclusion}.

\newpage
\section{State-dependent Poisson process} \label{sec:sdp}
\subsection{Equation of motion} \label{sec:eom}
The SDP process is a doubly stochastic renewal process which models the ``stress'' in the system as a function of time, $X(t)$, as
\begin{eqnarray}
{X(t) = X(0) + t - \sum_{i=0}^{N(t)}\Delta X_{i}}\,, \label{eq:xoft}
\end{eqnarray}
where $X$ and $t$ are expressed respectively in dimensionless units of $\xc$ (the critical stress in the system at which a stress-release event becomes certain), and $\tau$ (the time taken for the system to accumulate the critical stress $\xc$, in the absence of any stress-release events). We attach a physical interpretation to $X(t)$ in the solar flare context in Section~\ref{sec:map} and Appendix~\ref{app:toy}. The amount of stress released at the $i$-th event, $\Delta X_{i}$, is a random variable, drawn from a user-specified PDF, $\etadx$, where $X(t_i^-)$ is the stress in the system immediately prior to the $i$-th event. In the standard configuration, $\etadx$ is fixed as a power law, but other options exist \citep{Carlin2019quasi, Carlin2021endog}. Making $\eta$ conditional on $X(t_i^-)$ is necessary to ensure that the stress remains positive-definite and is plausible physically. The second random variable in Equation~\eqref{eq:xoft} is $N(t)$, a stochastic function which counts the number of events up to time $t$. It is determined iteratively via the waiting time between each event. 

We assume the instantaneous event rate is a monotonically increasing function of the stress in the system, viz.
\begin{eqnarray}
{\lambda[X(t)] = \frac{\alpha}{1 - X(t)}}\,, \label{eq:lam}
\end{eqnarray}
where $\alpha = \lambda_0 \tau$ is a dimensionless control parameter, and $\lambda_0 = \lambda(X=1/2)/2$ is a reference rate. The long-term statistical output of the SDP process does not depend strongly on the functional form of $\lambda[X(t)]$, so long as it diverges in the limit $X \rightarrow 1$ as in Equation~\eqref{eq:lam} \citep{Fulgenzi2017, Carlin2019quasi}. Equation~\eqref{eq:lam} implies that the probability of an event occurring approaches one, as the stress approaches the critical stress. As the stress increases deterministically between events, the PDF of waiting times $\Delta t$ following the $i$-th event is that of a variable-rate Poisson process \citep{Cox1955, Fulgenzi2017}
\begin{eqnarray}
p[\Delta t\,|X(t_i^+)] =&~ \lambda&[X(t_i^+) + \Delta t] \label{eq:pdelt} \\
 &\times& \exp \left\{ -\int_{t_i^+}^{t_i^+ + \Delta t}\td t' \lambda \left[ X(t') \right]\right\}\,, \nonumber
\end{eqnarray}
where $X(t_i^+)$ is the stress immediately following the event at time $t_i$.  

\subsection{Monte Carlo automaton} \label{sec:auto}
Analytically solving the coupled equations~\eqref{eq:xoft}--\eqref{eq:pdelt} to calculate the PDFs of event waiting times or sizes is usually not feasible, except for particular choices of $\etadx$; see section 6 of \citet{Fulgenzi2017} for an example. Instead, it is simple to run the following automaton to generate numerical solutions:
\begin{enumerate}
\item Pick $\Delta t$ from Equation~\eqref{eq:pdelt}, given the current stress $X$.
\item Update the stress to $X + \Delta t$ to account for the deterministic evolution.
\item Pick $\Delta X$ from $\eta\left[\Delta X \,|\, X + \Delta t \right]$, and subtract it from the stress.
\item Repeat from step 1.
\end{enumerate}
Given $\alpha$ and $\etadx$ the automaton generates a time-ordered sequence of waiting times and sizes. From the sequence we can calculate the long-term PDFs for waiting times and sizes \citep{Fulgenzi2017, Carlin2019quasi}, as well as the cross-correlation \citep{Melatos2018, Carlin2019quasi}, autocorrelations \citep{Carlin2019ac}, and other observables.

\begin{figure*}
\centering
\includegraphics[width=0.95\linewidth]{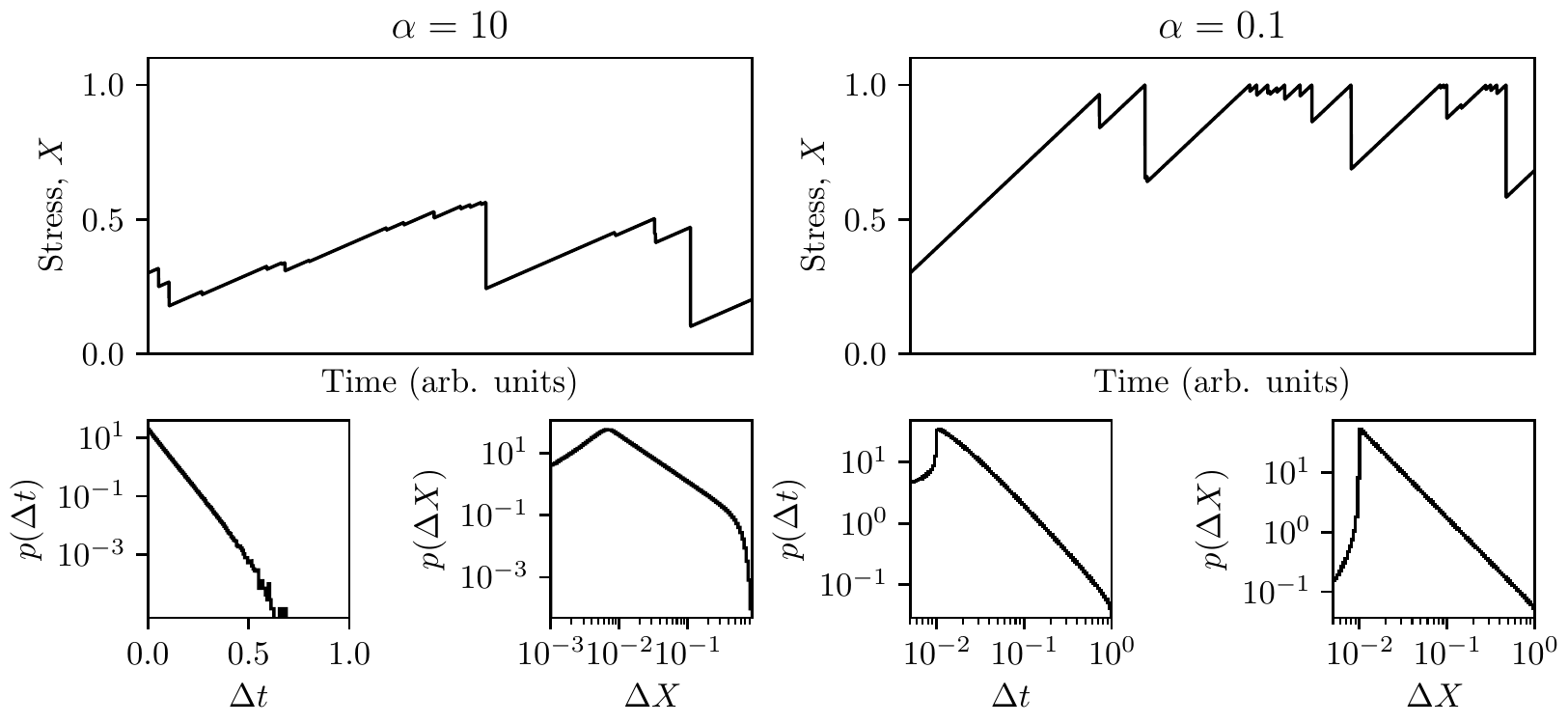}
\caption{Top panels: qualitative behavior of the evolution of the stress, $X(t)$, in the SDP process in the two main regimes, slowly driven (left panel, $\alpha=10$), and rapidly driven (right panel, $\alpha=0.1$). Bottom panels: predicted waiting time and size PDFs $p(\Delta t)$ and $p(\Delta X)$ respectively in the above regimes. For all panels we fix $\etadx \propto \left(\Delta X_{i}\right)^{-3/2}H\left[\Delta X_{i} - 10^{-2}X(t_i^-)\right]$, where the Heaviside function $H(...)$ enforces a minimum stress-release size of 1\% of the stress in the system, ensuring integrability. The histograms in the bottom panels each include $N=10^7$ events.}
\label{fig:sdp}
\end{figure*}

\subsection{Mapping to solar flares} \label{sec:map}
We identify the stress that accumulates between flares and relaxes at a flare with the spatially-averaged magnetic energy density in a given active region. One could equally choose a different physical quantity, such as the magnetic shear, depending on the particular microphysics of the flare trigger. We assume that active regions have independent stress reservoirs, i.e.~the coronal magnetic fields of different active regions do not interact strongly, and all flares from the same active region extract energy from one reservoir. An idealized toy model relating the foregoing definition of stress to magnetic energy input from the photosphere is outlined in Appendix \ref{app:toy}, as a simplified but concrete illustration of the physical picture under consideration. 

Two major simplifying assumptions are that the rate at which energy is fed into the reservoir, $\tau^{-1}$, is constant in time, and so is the critical, spatially-averaged magnetic energy density, $\xc$, for a given active region. These assumptions are motivated in Appendix \ref{app:toy}, but a key goal of the paper is to test their veracity. We also assume that flares reduce $X(t)$ instantaneously. This assumption is defensible as the typical time between flares (typically hours) is much larger than the typical duration of flaring events (typically minutes) \citep{Fletcher2011}. In what follows we do not prescribe a particular functional form for $\etadx$, to keep the SDP process as flexible as possible. 

The SDP process generates sequences of dimensionless waiting times, $\left\{\Delta t_{i}^{\rm SDP}\right\}$ and sizes, $\left\{\Delta X_{i}^{\rm SDP}\right\}$. It is natural to ask how they correspond to directly observable sequences of waiting times, $\left\{\Delta t_{i}^{\rm obs}\right\}$ and sizes, $\left\{\Delta s_{i}\right\}$. The waiting times satisfy
\begin{eqnarray}
{\Delta t_{i}^{\rm obs} = \Delta t_{i}^{\rm SDP} \tau}\,,\label{eq:delt_prop}
\end{eqnarray}
where $\tau$ is unknown a priori but can be estimated, as discussed in Appendix \ref{app:hierarch}. The sizes are related in a more complicated way, because some energy is released through channels other than the soft X-ray flux. We make the simplifying assumption that the peak soft X-ray flux multiplied by the duration of the flare (henceforth $\Delta s_{i}$ for the $i$-th flare in a given active region) is proportional to the stress released from the reservoir. This implies
\begin{eqnarray}
{\Delta s_{i} \propto \Delta X_{i}^{\rm SDP} \xc}\,.\label{eq:delx_prop}
\end{eqnarray}
The unknown constant of proportionality in Equation~\eqref{eq:delx_prop} hampers direct parameter estimation of $\xc$ from an observed size PDF. 

Full parameter estimation for the SDP process given a set of observed waiting times and sizes is discussed in \citet{Melatos2019}. It lies outside the scope of this paper. In Appendix \ref{app:hierarch} we also describe for completeness an alternative parameter estimation procedure using a hierarchical Bayesian scheme. The advantage of the hierarchical scheme is that it clarifies as a matter of principle the information content and flow in the estimation problem, i.e.~the parameter combinations that can be estimated uniquely, and the data components that inform each parameter estimate. The disadvantage is that it can be implemented in practice only if $\etadx$ is drawn from a certain class of mathematical functions. The favored class produces waiting time and size PDFs which do not match solar flare observations (see Section~\ref{sec:disagg_general}), so the hierarchical scheme is not applied to real data in this paper. Nonetheless, it is included for the benefit of the reader, who may wish to develop it further for solar flare analysis in the future.

\section{Observable signatures of a rapidly-driven process} \label{sec:pred}
Broadly speaking, an SDP process operates in one of two regimes: slowly-driven, with $\alpha \gg 1$, and rapidly-driven, with $\alpha \ll 1$. In this section we identify the key dynamics in both regimes, in Sections~\ref{sec:slow} and \ref{sec:fast} respectively. We then infer, in Section~\ref{sec:heur}, a set of observable signatures, which if absent rule out the operation of a rapidly-driven process of the form described in Section~\ref{sec:fast}.

\begin{figure}
\centering
\includegraphics[width=0.95\linewidth]{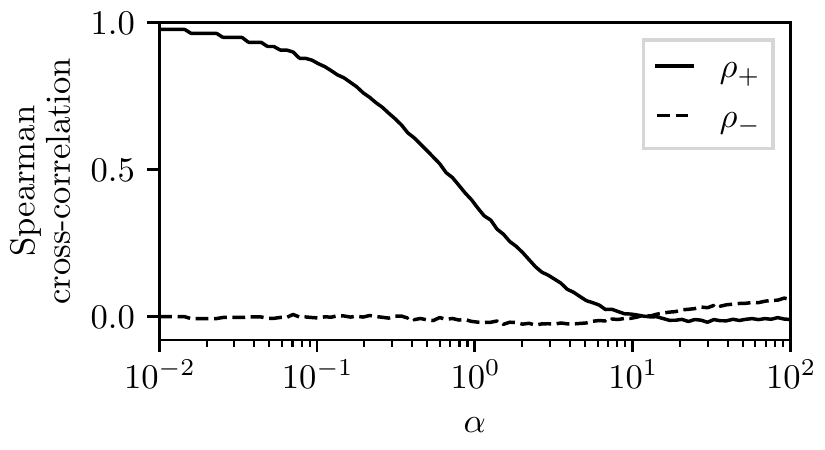}
\caption{Forward, $\rho_+$ (solid curve), and backward, $\rho_-$ (dashed curve), cross-correlations between event sizes and waiting times for $10^2$ values of $10^{-2} \leq \alpha \leq 10^2$, with $N=10^5$ events generated per value of $\alpha$. We fix $\etadx$ to the same functional form as in Figure~\ref{fig:sdp}. A version of this figure first appeared as figure 13 in \citet{Fulgenzi2017}.}
\label{fig:sdp_correl}
\end{figure}

\subsection{Slow driver} \label{sec:slow}
When the system is slowly driven, we have $X(t) \ll 1$, as events are usually triggered before the stress accumulates to near the threshold. When $\etadx$ is a power law, the predicted waiting time PDF $p(\Delta t)$ is an exponential, and the predicted size PDF $p(\Delta X)$ is a power law over multiple decades. We show the qualitative behavior of the stress in the slowly driven regime in the top-left panel of Figure~\ref{fig:sdp}, where for the sequence of 20 stress-release events shown we have $0.1 < X(t) < 0.6$. In the bottom-left panels of the same figure we show the predicted waiting time and size PDFs. The size PDF $p(\Delta X)$ has a turn-over in logarithmic slope at $\Delta X \approx 10^{-2}$ due to the particular choice of $\etadx$ (see caption for the functional form). 

Depending on the choice of $\etadx$, there may be detectable backward cross-correlations (i.e.~a correlation between the size of an event and the waiting time since the preceding event). This is because the size of an event cannot exceed the amount of stress in the system, so longer periods of stress-accumulation allow for the possibility of larger events. This backward cross-correlation is especially pronounced with the choice $\etadx \propto \delta\left[\Delta X - X(t_i^-)\right]$, where $\delta(\dots)$ is the Dirac-delta function. This choice forces the stress reservoir to empty at each event, and collapses the SDP process to other stochastic processes in the literature, e.g.~forest fire models \citep{Daly2006}. In the solar flare context, a backward cross-correlation is a long-standing prediction of the \citet{Rosner1978} stress build-up model; and the ``reset'' model of \citet{Hudson1998, Hudson2019}. If instead $\etadx$ prefers small stress-release events, e.g.~if it is a power law \citep{Fulgenzi2017}, the backward cross-correlation is small. 

\subsection{Fast driver} \label{sec:fast}
When the system is rapidly driven, we have $X(t^-_i) \lesssim 1$, i.e.~the stress is driven close to the threshold before each event. As the dynamics of the system are strongly influenced by the presence of the threshold, we sometimes call this regime ``threshold-driven''. A prediction of the SDP process in this regime is that the size and waiting time PDFs should ``match'', i.e.~they should be the same distribution, up to a linear scaling \citep{Fulgenzi2017}. For example, if $\etadx$ is a power law, both the waiting time and size PDFs are power laws. We show this in the bottom-right panels of Figure~\ref{fig:sdp}. We also show the qualitative behavior of the stress versus time in this regime in the top-right panel of the same figure, where we see $X(t) \rightarrow 1$ (i.e.~unit critical stress) before every stress-release event.

A strong forward cross-correlation (i.e.~a correlation between the size of an event and the waiting time to the next event) is observed in this regime. This is because a large stress-release event results in a longer delay before the system accumulates enough stress to approach the threshold. In the solar flares context, this cross-correlation is predicted by the ``saturation'' model of \citet{Hudson1998} and \citet{Hudson2019}.  The smooth transition between the fast-driven regime ($\alpha \ll 1$), with high forward cross-correlations, and the slow-driven regime ($\alpha \gg 1$), is shown in Figure~\ref{fig:sdp_correl}.

\subsection{Heuristic test for a threshold-driven process} \label{sec:heur}
Ideally, we fit the parameters of the SDP process ($\tau$, $\xc$, $\lambda_0$, and $\etadx$) to a paired sequence of solar flare sizes and waiting times, and infer what regime applies. In practice, however, at least some of these parameters vary between active regions, which limits us to sequences of length $N \lesssim 50$ (typical active regions have at most dozens of events; see Section~\ref{sec:data}), which are insufficient to infer four or more parameters. Instead, we combine the results in Sections~\ref{sec:slow} and \ref{sec:fast} to deduce qualitative features, which must be present in multiple observables simultaneously (e.g.~PDFs, cross-correlations), if the SDP process operates in a particular $\alpha$ regime or indeed operates at all. Specifically, we summarize the behavior in Sections~\ref{sec:slow} and \ref{sec:fast} into the following observationally testable prediction. If solar flares are well-modeled with a rapidly-driven process (e.g.~the SDP process with $\alpha \ll 1$), then we should see large forward cross-correlations, accompanied by waiting time and size PDFs with the same shape in individual active regions. This coincident signature should become more prominent as $\alpha$ (or a proxy thereof) decreases. We quantify this prediction in Section~\ref{sec:corr}.

With the test above we also ameliorate the impact of a potentially mis-specified model. Although it is agnostic about the microphysics, the SDP process is not the only possible prescription of stress accumulation and release. If some of the assumptions outlined in Section~\ref{sec:map} do not hold, a different underlying model may underpin solar flares. For example, one may build a model in which stress accumulates according to a Brownian random walk with some underlying drift, until a threshold is breached, at which point a stress-release event is triggered \citep{Carlin2020bsa}. In the Brownian stress accumulation model, $X(t)$ is always driven to the threshold before each event. If drift occurs faster than diffusion, we would predict the same coincident signature as for the SDP process: large forward cross-correlations are accompanied by matching waiting time and size PDFs.

\section{\emph{GOES} soft X-ray observations} \label{sec:data}
The X-ray Sensor (XRS) on the \emph{Geostationary Operational Environmental Satellites} (\emph{GOES}) has continuously monitored the Sun in soft X-rays (0.5 to 8\,\AA) since 1975. In Section~\ref{sec:summary} we introduce the flare summary data analyzed in this paper. In Section~\ref{sec:obscuration} we remind the reader of the obscuration effect described in \citet{Wheatland2001}, explain how it affects our analysis, and briefly touch on other biases the \emph{GOES} flare detection algorithm may have on the completeness of the catalog as a whole. In Section~\ref{sec:agg} we analyze aggregated waiting time and size PDFs across all active regions. 

\subsection{Flare data} \label{sec:summary}
We use the publicly available flare summary data hosted by the National Geophysical Data Center (NGDC)\footnote{\url{ftp://ftp.ngdc.noaa.gov/STP/space-weather/solar-data/solar-features/solar-flares/x-rays/GOES/xrs/}} for flares before June 29 2015, and by the Space Weather Prediction Center (SWPC)\footnote{\url{ftp://ftp.swpc.noaa.gov/pub/indices/events/}} for flares after June 28 2015. We combine these two data sources into a homogeneous, cleaned database (henceforth ``catalog''), with some anomalies corrected as described in Appendix \ref{app:cleaning}. These data are collated with the flare start epochs, $t^{\rm s}$, peak epochs, $t^{\rm p}$, and end epochs, $t^{\rm e}$. The peak epoch is defined as the epoch which contains the highest peak flux, after the start epoch. The end epoch is defined as the epoch at which the flux returns to half of the difference between the peak flux and the background flux. The peak flux (irradiance) of each flare, $f^{\rm p}$, is calculated with reference to the recorded flare class, in units of W\,m$^{-2}$. The longitude and latitude of the flare are also often available, if the flare is associated with an active region. For clarity, we henceforth denote as $x_{i,\,k}$ an arbitrary measurement of the variable $x$ in the $i$-th flare from the $k$-th active region.

We define the waiting time between two flare epochs as $\Delta t_{i,\,k} = t^{\rm p}_{i+1,\,k} - t^{\rm p}_{i,\,k}$. One could equally use the flare start or end epochs to define the waiting time. We use the flare peak epoch as it is less influenced by the background level, as discussed more in Section~\ref{sec:obscuration}. We define the flare size as $\Delta s_{i,\,k} = f^{\rm p}_{i,\,k} \left(t^{\rm e}_{i,\,k} - t^{\rm s}_{i,\,k}\right)$; that is, we multiply the irradiance by the duration of the flare to obtain a flare size with dimensions of energy per unit area. The duration of an active region is $\Delta T_{k} = t^{\rm e}_{N_k,\,k} - t^{\rm s}_{1,\,k}$, where there are $N_k$ flares in the region. The average flare rate in an active region is $\lambda_k = N_k / \Delta T_k$. The average $\lambda_k$ overestimates systematically the true flare rate, as the duration $\Delta T_k$ is between two flare epochs, both of which are counted in $N_k$. An unbiased estimate would take $\Delta T_k$ as the difference between the epochs when the active region appears and disappears, but neither epoch is recorded in the \emph{GOES} flare summary data\footnote{The epochs when the active region appears and disappears may be extracted from other records in some instances. Such an analysis lies outside the scope of this paper.}. A concern we touch on further in Section~\ref{sec:agg} is that $\lambda_k = N_k / \Delta T_k$ assumes that the flare catalog is complete, i.e.~all flares from a given region are detected and correctly attributed to that region.

\begin{figure}
\centering
\includegraphics[width=0.95\linewidth]{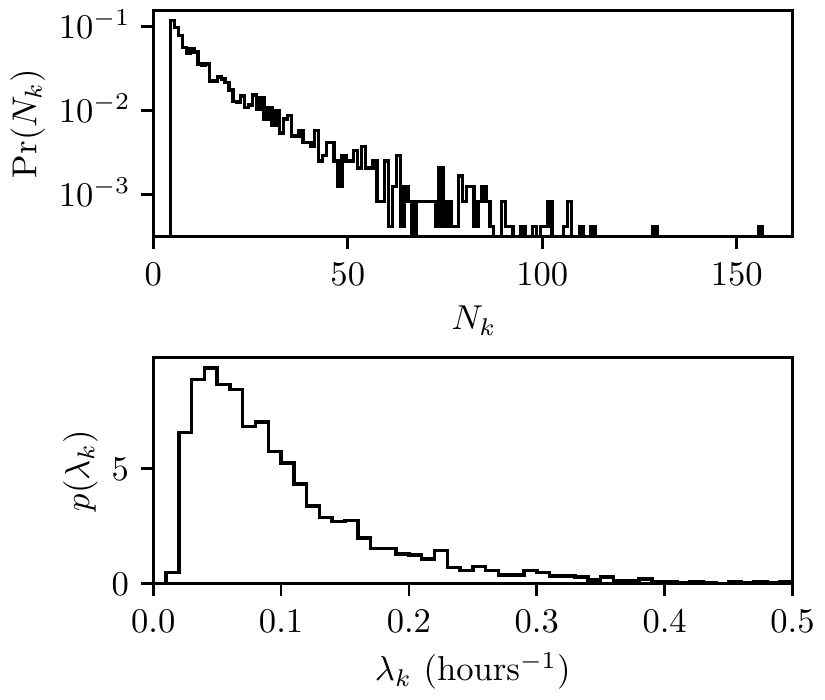}
\caption{Top panel: Probability mass function of the number of flares per active region, $N_k$, for all active regions with $N_k \geq 5$. Regions with $0 \leq N_k \leq 4$ are excluded, as the number of samples per region is too small for the statistical tests in this paper. Bottom panel: PDF of the flare rate, $\lambda_k$, for all active regions with $N_k \geq 5$, binned into 50 uniformly spaced bins between $0\,\textrm{hours}^{-1}$ and $0.5\,\textrm{hours}^{-1}$.}
\label{fig:hist_numrates}
\end{figure}

As of September 1 2022, there are 83283 flares in the catalog, 49328 of which are associated with an active region. There are 2429 active regions with $N_k\geq 5$, accounting for 42982 flares. The median number of flares in an active region (when considering only regions with $N_k \geq 5$) is 12, but the average is 17.7, as the distribution is peaked at $N_k = 5$ and monotonically decreases with $N_k$. We show the probability mass function of the number of flares in an active region, ${\rm Pr}(N_k)$, in the top panel of Figure~\ref{fig:hist_numrates} for regions with $N_k \geq 5$. The PDF of observed flare rates, $p(\lambda_k)$, across all active regions with $N_k \geq 5$, is displayed in the bottom panel of the same figure. The distribution of $\lambda_k$ is well-described by a log-normal, with mean $0.08\,$hours$^{-1}$, and standard deviation $0.7\,$hours$^{-1}$. Henceforth, we typically only include regions with $N_k \geq 5$ in our analysis unless stated otherwise, as many of the subsequent statistical tests, such as the cross-correlation(s), have low statistical power with smaller sample sizes.

\subsection{Obscuration and flare size bias of subsequent flares after a large flare} \label{sec:obscuration}
As described in section 2.2 of \citet{Wheatland2001}, the \emph{GOES} flare detection algorithm involves a selection effect that obscures the detection of flares following a large flare, due to the enhanced background soft X-ray flux. To detect a flare, the flux must monotonically increase for four consecutive minutes, with the last value 1.4 times the value three minutes earlier. Hence a flare must produce a 40\% increase above the background flux. However, while flares typically rise rapidly to peak flux, the soft X-ray emission is observed to decay on the timescales of hours \citep{Benz2016}. Thus, even large flares may be obscured by the enhanced background flux following, say, an X1 ($f^{\rm p} = 10^{-4}\textrm{\,W\,m}^{-2}$) class flare. 

The above detection algorithm introduces a secondary bias in the catalog. A flare, say with recorded peak flux $f^{\rm p}_i$, that occurs during the decay of another flare, say with recorded peak flux $f^{\rm p}_{i-1}$, has an enhanced peak flux compared to if the same flare occurs during a period of low background flux, viz.  
\begin{equation}
f^{\rm p}_i \approx f^{\rm p}_{i,\,\textrm{true}} + f^{\rm p}_{i-1}\exp\left(\frac{-\Delta t_{i-1}}{\tau}\right)\,,
\end{equation}
where $\tau$ is the timescale on which the peak flux from the $(i-1)\,$-th flare decays, and $f^{\rm p}_{i,\,\textrm{true}}$ is the ``true'' peak flux of the $i$-th flare. A secondary effect is that the duration of the $i$-th flare will have a reduced duration due to the decaying contribution of the previous flare to the background flux. That is, during the $i$-th flare, the background flux decays by a factor of $\exp\left[-\left(t^{\rm e}_i - t^{\rm s}_i\right) / \tau\right]$, which reduces the time taken for the flux to decay back to half the difference between the peak and pre-flare fluxes. These two effects run counter to one another, as $\Delta s_i$ is defined as the product of peak flux and duration. They are likely present in a small subset of the catalog, as only $1.2\%$ of all flares from regions with $N_k \geq 5$ have start times before the end time of the previous flare.

A comprehensive re-analysis of the soft X-ray flux time-series, with appropriate background subtraction, may reveal flares that were not detected with the original flare detection algorithm and correct the biases in flare size outlined above. Such a re-analysis lies outside the scope of this paper. Acknowledging that the \emph{GOES} catalog is incomplete, we mitigate the impact on our analysis by creating a secondary masked catalog that only includes flares with $\Delta s \geq 10^{-3}\textrm{\,J\,m}^{-2}$ (i.e.~class C1 multiplied by the median duration of $10^3\,$s, and higher), where we know the fraction of missed flares is smaller. This masked catalog is used in Sections~\ref{sec:agg} and \ref{sec:disagg} when we perform comparative studies and parametric fits for the waiting time and size PDFs. 

\subsection{Aggregate size and waiting time statistics} \label{sec:agg}
\begin{figure}
\centering
\includegraphics[width=0.95\linewidth]{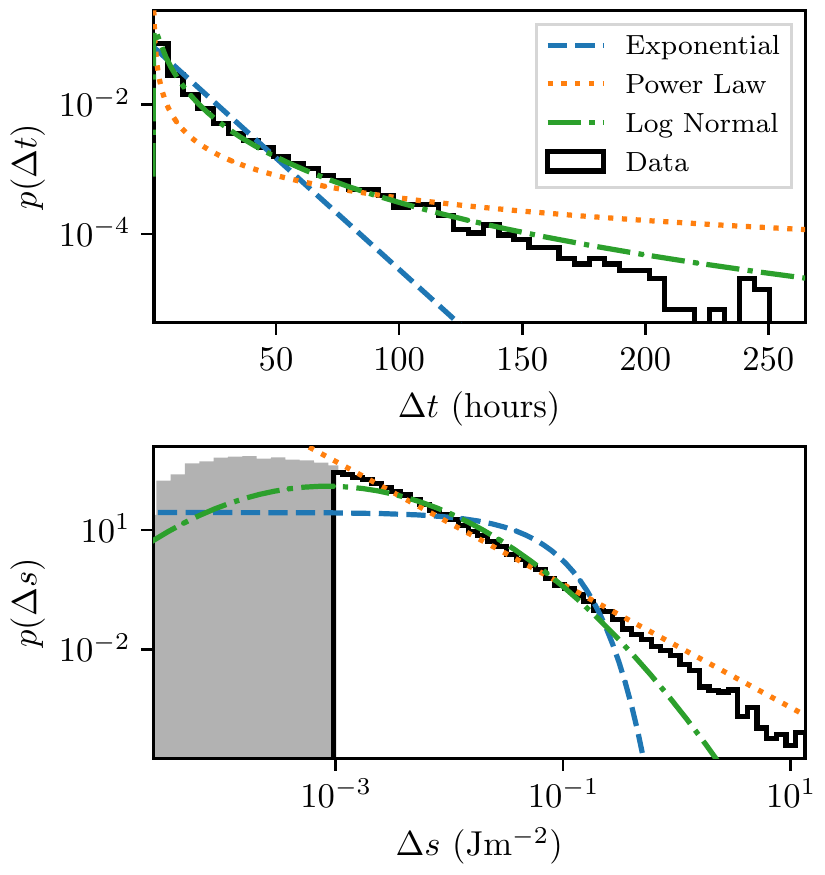}
\caption{Top panel: PDF $p(\Delta t)$ of waiting times (thick black stepped curve), $\Delta t$, aggregated over all active regions with $N_k \geq 5$, binned into 50 linearly spaced bins between the minimum and maximum $\Delta t$ in the full, unmasked catalog. Overlaid are the best fit estimates of an exponential (blue dashed curve), power law (orange dotted curve), and log-normal (green dot-dashed curve) PDF. Bottom panel: As for top panel, but for sizes $\Delta s$, and the data is binned into 50 logarithmically spaced bins. The grey region in the bottom panel shows the difference between the size PDFs from the masked and full catalogs (see Section~\ref{sec:obscuration}).}
\label{fig:pdfs_overall}
\end{figure}

When we aggregate flare waiting times and sizes across all active regions with $N_k \geq 5$, we obtain the PDFs shown in the top and bottom panels of Figure~\ref{fig:pdfs_overall} respectively. The data are shown as the black histograms. The grey region in the bottom panel shows the flare sizes that are not included in the masked catalog, as described in Section~\ref{sec:obscuration}. The difference between the masked and full catalog is not visible in the top panel as the effect on $p(\Delta t)$ is minimal. 

Three empirical trial distributions with common analytic forms are overlaid on each PDF for the masked catalog. The overlaid distributions have parameters fixed to their maximum likelihood values, given the data. By eye, it is clear that the log-normal distribution best describes the aggregated waiting time PDF\footnote{Waiting times that are best described with a log-normal are seen in many contexts. We provide a brief survey in Appendix~\ref{app:ln} for the interested reader.}, $p(\Delta t)$, while a power law distribution best describes the aggregated size PDF, $p(\Delta s)$. We formalize this comparison using the corrected Akaike Information Criterion (AICc) \citep{Akaike1974, Hurvich1989}. The AICc calculates the model, among a set of possible models, which minimizes the information loss, while accounting for potential bias due to the number of model parameters and the sample size. When calculated for $p(\Delta t)$, the relative probability for a log-normal describing the data over an exponential or a power law is $e^{3\times10^3}$ and $e^{3\times10^4}$ respectively. When calculated for $p(\Delta s)$, the relative probability for a power law describing the data over an exponential or a log-normal is $e^{3\times10^4}$ and $e^{4\times10^3}$ respectively. If we instead only mask flares with peak flux less than $10^{-6}\,$Wm$^{-2}$ (i.e.~class C1), we find that a log-normal best describes the data. This preference is also noted in \citet{Verbeeck2019}. It arises because of the small number of flares in this alternatively-masked catalog with $10^{-4} < \Delta s /\textrm{\,J\,m}^{-2} \lesssim 10^{-3}$. If we fit the full, unmasked catalog (i.e.~include the flares shaded in grey in the bottom panel of Figure~\ref{fig:pdfs_overall}) we find that $p(\Delta s)$ is again fitted best with a log-normal. We remind the reader that $p(\Delta s)$ is an observed distribution; it is a function of both the underlying generative physics (i.e.~how much energy is released in each flare) and the systematic observational biases (i.e.~how many and what flares are detected or not). The masking described in Section~\ref{sec:obscuration} is a first pass at accounting for some of these biases.

If we assume that the same $\alpha$ and $\etadx$ apply to all regions, Figure~\ref{fig:pdfs_overall} is inconsistent with the SDP framework for any choice of $\alpha$ and $\etadx$. However, as we discuss in Section~\ref{sec:map}, there is good reason to believe that $\alpha$ (and perhaps even $\etadx$) may vary region-to-region.

\newpage
\section{Disaggregated data in individual active regions} \label{sec:disagg_general}
The goal of this section is to search for signatures of a threshold-driven SDP process in individual active regions, rather than considering all flares in aggregate. In Section~\ref{sec:disagg} we consider only waiting time and size PDFs, without regard to their potential cross-correlation. In Section~\ref{sec:corr} we calculate these cross-correlations, and in Section~\ref{sec:matching} we apply the test outlined in Section~\ref{sec:heur} by searching for an association between matching waiting time and size PDFs and the cross-correlation, in individual active regions. In Section~\ref{sec:ac} we perform a preliminary investigation of the longer-term memory in the system by calculating the autocorrelation between subsequent waiting times and subsequent sizes. The analysis in Section~\ref{sec:disagg} involves parametric fitting of the size PDFs, so we use the masked catalog described in Section~\ref{sec:obscuration}. However in Sections~\ref{sec:corr}--\ref{sec:ac} we use the full, unmasked catalog, as the tests performed in these sections are non-parametric.

\subsection{Waiting time and size PDFs versus flare rate} \label{sec:disagg}
For regions with $N_k \geq 5$ we disaggregate the data, and ask what PDF shape best characterizes each region's waiting time and size PDFs, rather than considering only the aggregated dataset as in Section~\ref{sec:agg}. Using the AICc, we find that for waiting time PDFs, 63\% of regions are fitted best with an exponential distribution, 20\% with a log-normal, and the remainder with a power law. For size PDFs, 63\% of regions are fitted best with a power law distribution, 24\% with an exponential, and the remainder with a log-normal. These proportions are broadly consistent with previously published results, i.e.~that while the aggregated $p(\Delta t)$ is fitted best with a log-normal, individual regions are often fitted best with an exponential (with varying rates) \citep{Wheatland2000delt}. 
Individual regions have $p(\Delta s)$ that is usually fitted best with a power law, but can occasionally be better represented with log-normal or exponential distributions, especially in regions with lower $N_k$. As an example of the different shapes that distributions of waiting times and sizes can have in different regions, we display four arbitrary but representative per-active-region fits in Appendix~\ref{app:perar}.

\begin{figure}
\centering
\includegraphics[width=0.95\linewidth]{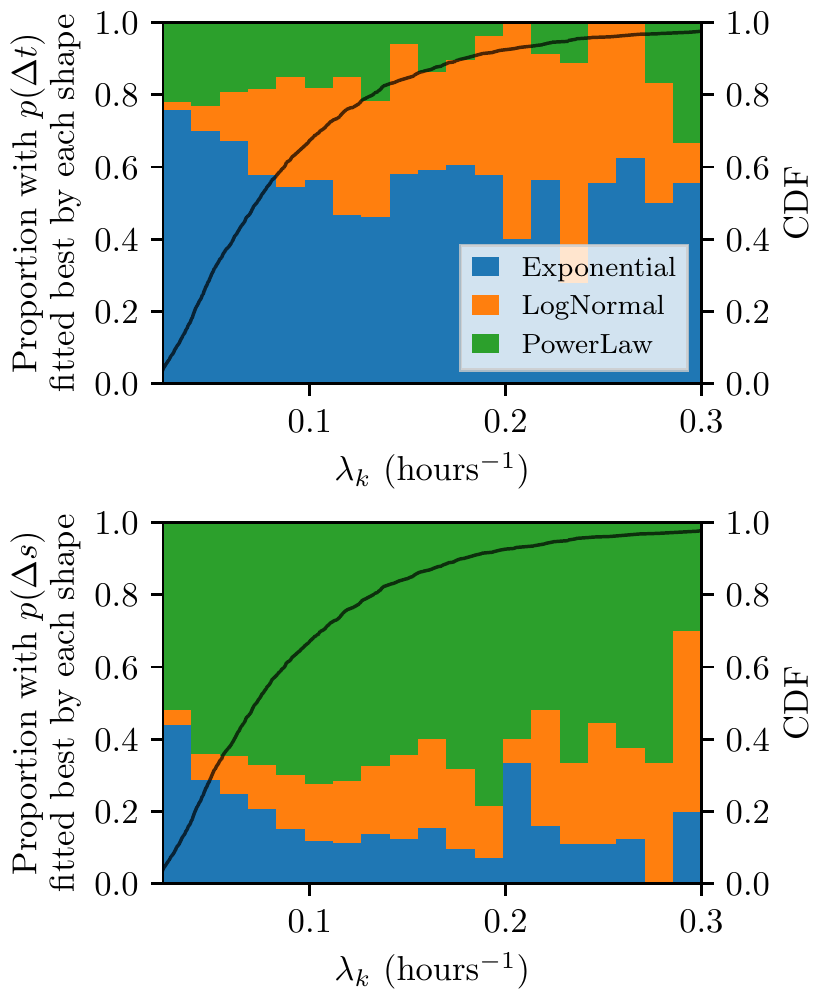}
\caption{Top panel: Proportion of active regions with waiting time PDF fitted best by three possible shapes (a power law, an exponential, and a log-normal) as determined by the AICc, as a function of the region's flare rate, $\lambda_k$. The CDF of $\lambda_k$ is overlaid in black. Bottom panel: As in the top panel, but for size PDFs.}
\label{fig:prop_both}
\end{figure}

One may reasonably ask whether there are clear trends, or predictors, for the waiting time and size PDFs in any given region. An exhaustive search for predictors, e.g.~with a multiple regression analysis, lies outside the scope of this paper, but one sensible first step is to see if the shape that fits best evolves with $\lambda_k$. This test is performed for the waiting time and size PDFs in the top and bottom panels of Figure~\ref{fig:prop_both} respectively. This figure is constructed by binning all active regions with $N_k \geq 5$ into 20 uniformly spaced bins between $\lambda_k = 0.025\,\textrm{hours}^{-1}$ and $\lambda_k = 0.3\,\textrm{hours}^{-1}$, then calculating which distribution fits best the waiting times and sizes using the AICc. For the waiting time PDFs, the proportion of regions fitted best with a power law stays roughly constant as $\lambda_k$ increases, at around 15\%, while the proportion of regions fitted best with a log-normal grows with $\lambda_k$, at the expense of the exponential. For the size PDFs we see a similar trend, with the proportion of regions fitted best with a power law staying roughly constant as $\lambda_k$ increases, this time at around 65\%, while the proportion of regions fitted best with a log-normal grows with $\lambda_k$, at the expense of the exponential. In both panels we plot the cumulative distribution function (CDF) of $\lambda_k$ to remind the reader that each $\lambda_k$ bin does not contain the same number of regions; 90\% of regions with $N_k \geq 5$ have $0.028 < \lambda_k / (1\,\textrm{hour}^{-1}) < 0.27$.

Can the evolution of the proportion of each shape versus $\lambda_k$ be explained with the SDP model? Under the assumption that $\lambda_k$ is a tracer of the driving rate, i.e.~$\lambda_k \propto \alpha^{-1}$, we expect to see a greater proportion of exponentially distributed waiting times at low $\lambda_k$ (high $\alpha$). This is the regime in which the stress does not approach the threshold at $X = \xc$ before each event, so waiting times are un-correlated with sizes and are (broadly) Poissonian, i.e.~the waiting times are exponentially distributed. This expectation broadly conforms with what we see in the top panel of Figure~\ref{fig:prop_both}. 

The evolution versus $\lambda_k$ in the bottom panel of Figure~\ref{fig:prop_both} is harder to explain with the SDP model, if each region has the same $\etadx$. We expect $p(\Delta s) \propto \eta$, when $\alpha$ is low, but we see all of the three possible shapes represented at the highest values of $\lambda_k$. This implies at least one of the following: \begin{enumerate*}[i)] \item $\eta$ truly varies from one region to the next, which implies different stress-release mechanisms are at play in different regions; or \item $\lambda_k \gtrsim 0.2\,$hours$^{-1}$ does not correspond to $\alpha \ll 1$, and hence $p(\Delta s) \propto \eta$; or \item the small sample size of events in each region results in the AICc not favoring the ``true'' size distribution; or \item the obscuration effects described in Section~\ref{sec:obscuration} is stronger with higher $\lambda_k$, due to the enhanced background flux in regions that have many flares in a short period of time. \end{enumerate*} To test iii), we generate Figure~\ref{fig:prop_both} again for regions with $N_k \geq 10$, instead of $N_k \geq 5$. For both the waiting times and sizes, the proportion fitted best by an exponential drops by $\sim 10\%$ in each $\lambda_k$ bin, while the proportion fitted best by a log-normal increases. Qualitatively, however, the evolution with $\lambda_k$ remains consistent with what is seen in Figure~\ref{fig:prop_both}.

\subsection{Size--waiting-time cross-correlations} \label{sec:corr}
The correlation between flare sizes and subsequent waiting times, i.e.~the ``forward'' cross-correlation, is denoted as $\rho_{+,\,k}$, while the correlation between flare sizes and the preceding waiting times, i.e.~the ``backward'' cross-correlation, is denoted as $\rho_{-,\,k}$. We calculate these correlations using the Spearman correlation coefficient \citep{Lehmann2006}.

\begin{figure}
\centering
\includegraphics[width=0.95\linewidth]{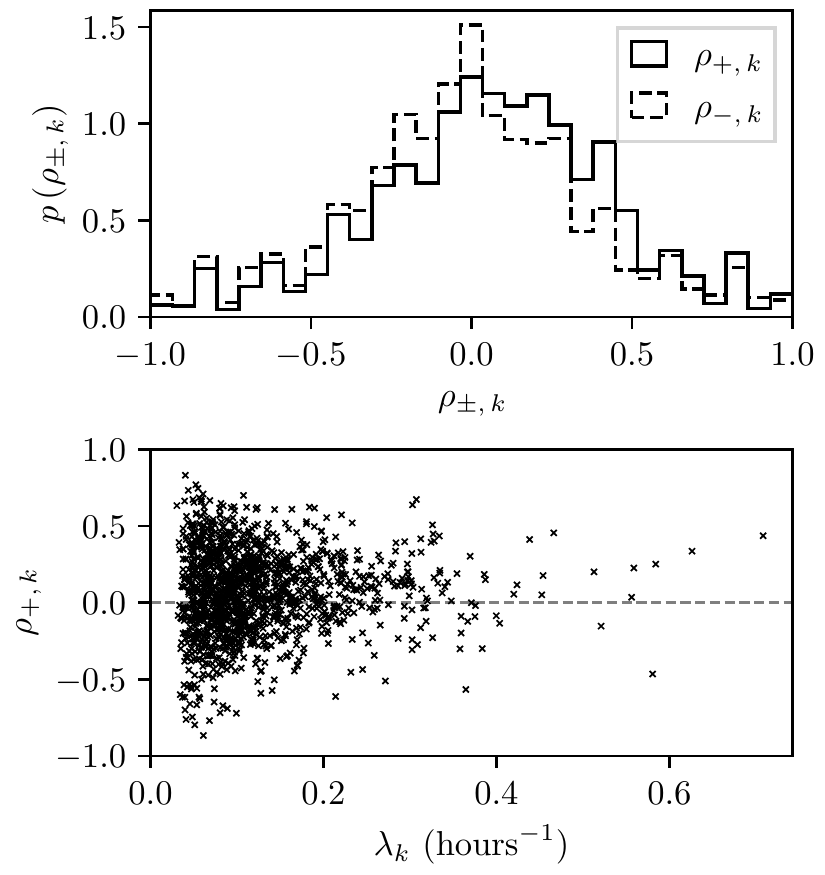}
\caption{Top panel: PDF of forward cross-correlation, $\rho_{+,\,k}$ (black stepped curve), and backward cross-correlation, $\rho_{-,\,k}$ (black dashed curve), for all active regions with $N_k \geq 5$. Data binned into 30 uniformly spaced bins between $\rho_{\pm,\,k} = -1$ and $\rho_{\pm,\,k} = 1$. Bottom panel: Scatter plot of the forward cross-correlation, $\rho_{+,\,k}$, and flare rate, $\lambda_k$, for all regions with $N_k \geq 10$. The dashed grey line corresponds to $\rho_{+,\,k}=0$.}
\label{fig:hist_corr}
\end{figure}

The PDFs of forward and backward cross-correlations measured in all active regions with $N_k \geq 5$ are displayed in the top panel of Figure~\ref{fig:hist_corr}. The PDFs are broad, mostly because the median $N_k$ is 12 (i.e.~small). However, $p(\rho_{+,\,k})$ and $p(\rho_{-,\,k})$ are different; the latter has a median of $\rho_{-,\,k} = 0.0$, while the former has a median of $\rho_{+,\,k} = 0.08$. A Kolmogorov-Smirnov (KS) two-sample test provides quantitative evidence of the difference, returning a $p$-value of $10^{-14}$.

The SDP process predicts a high forward cross-correlation for $\alpha \ll 1$, as discussed in Section~\ref{sec:fast}. The data do not show evidence in favor of such a correlation in the majority of active regions. In the bottom panel of Figure~\ref{fig:hist_corr} we do not see a clear visual trend between $\rho_{+,\,k}$ and $\lambda_k$ (for regions with $N_k \geq 10$), although a Spearman correlation test returns a small but non-zero correlation of $5\times10^{-2}$ ($p$-value of $0.06$). We remind the reader that both $\rho_{+,\,k}$ and $\lambda_k$ are empirical estimates for each active region, and the number of events per region is small. We do not plot uncertainties in the bottom panel of Figure~\ref{fig:hist_corr}, but they are typically large, of order the magnitude of the central estimate. These results imply that either \begin{enumerate*}[i)]
\item all active regions have $\alpha \gtrsim 1$, where no forward cross-correlation is expected, i.e.~$\lambda_k \gtrsim 0.2\textrm{\,hours}^{-1}$ does not correspond to the threshold-limited regime; or
\item some active regions have $\alpha \ll 1$, but either the trigger threshold or the driving rate is not constant with time, i.e.~the random process triggering flares does not conform to the assumptions made in the SDP framework.
\end{enumerate*}

As described in Section~\ref{sec:obscuration} the \emph{GOES} catalog is not complete, i.e.~while a given active region is visible, not all flares that occur are recorded. \citet{Wheatland2000corr} noted that the obscuration in Section~\ref{sec:obscuration} creates an artificial forward cross-correlation, which explains why the median $\rho_{+,\,k}$ is positive. 

\subsection{Matching the shapes of $p(\Delta t)$ and $p(\Delta s)$} \label{sec:matching}
\begin{figure}
\centering
\includegraphics[width=0.95\linewidth]{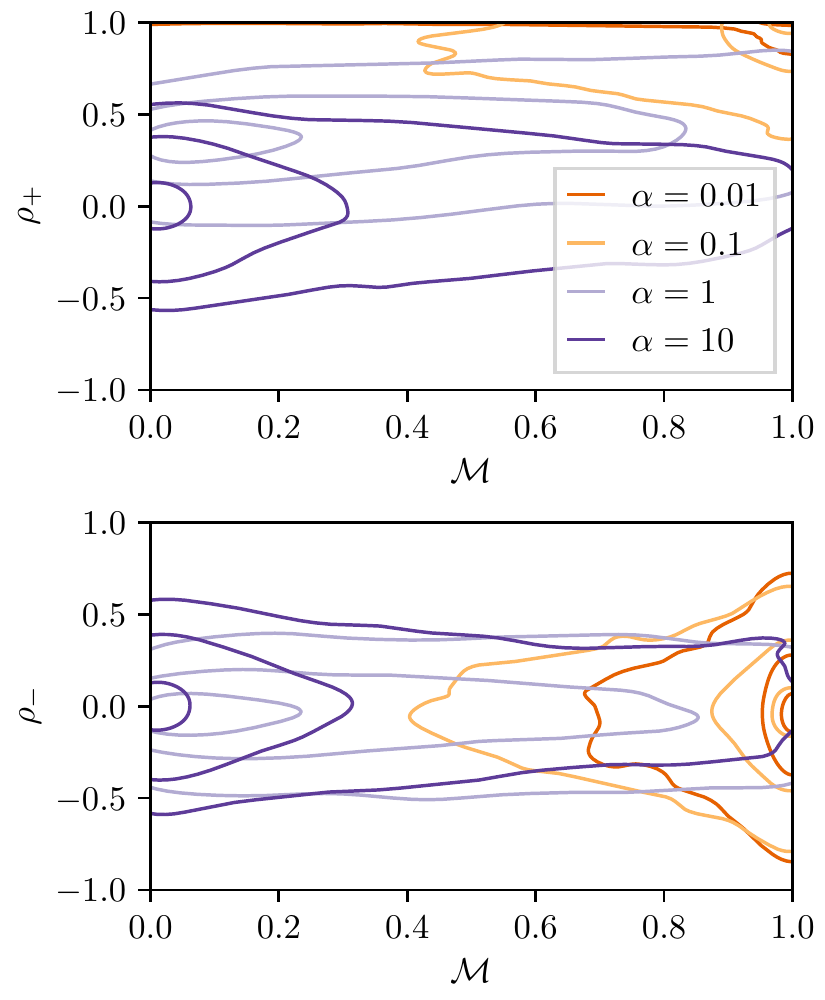}
\caption{Calibrating the cross-correlation test in Section~\ref{sec:matching}: two-dimensional KDEs of the relationship between $\rho_+$ and $\mathcal{M}$ (top panel), and $\rho_-$ and $\mathcal{M}$ (bottom panel), for events simulated from the SDP model with four values of $\alpha$ (see legend for color code). The text in Section~\ref{sec:matching} reports details about the Monte Carlo procedure. Contours correspond to the 10\%, 50\%, and 90\% credible intervals.}
\label{fig:sdp_kde}
\end{figure}

\begin{figure}
\centering
\includegraphics[width=0.95\linewidth]{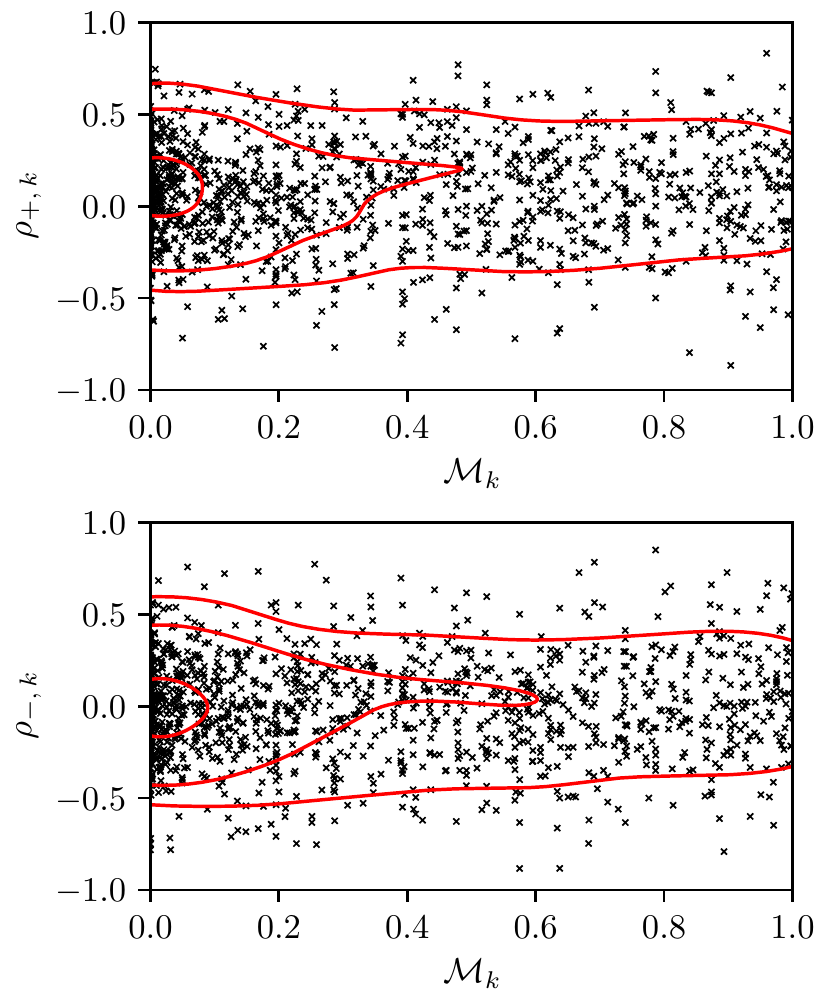}
\caption{As for Figure~\ref{fig:sdp_kde} but for \emph{GOES} data with black crosses marking all regions in the full, unmasked \emph{GOES} catalog with $N_k \geq 10$. Red contours correspond to the 10\%, 50\%, and 90\% credible intervals inferred from the black crosses.}
\label{fig:goes_kde}
\end{figure}

Another way to probe these data is to ask whether regions with relatively high $\rho_{+,\,k}$ exhibit ``matching'' functional forms for $p(\Delta t)$ and $p(\Delta s)$, as the SDP model predicts for $\alpha \lesssim 1$. We quantify the degree to which the PDFs match via the $p$-value $\mathcal{M}_k$ of the KS two-sample test applied to the sampled PDFs, $\{\Delta t_{i,\,k}\}$ and $\{\Delta s_{i,\,k}\}$. To calibrate, we perform a Monte Carlo simulation using sequences of events drawn from the SDP model. For each value of $\alpha$ we simulate $10^5$ ``regions''. For each region we generate $M$ events, where $M$ is a random number drawn from $p(N_k)$, restricted to values $N_k \geq 10$, as empirically measured for the \emph{GOES} regions (the distribution for $N_k \geq 5$ shown in Figure~\ref{fig:hist_numrates}). From these $M$ events we calculate summary statistics, e.g.~$\mathcal{M}$ and $\rho_+$. This results in $10^5$ pairs of $(\mathcal{M}, \rho_{+})$, from which we construct a two-dimensional kernel density estimate (KDE) \citep{Wand1995}. The KDE estimates the true joint PDF $p(\mathcal{M}, \rho_{+})$, and is qualitatively equivalent to a smoothed, two-dimensional histogram.

The result of this procedure for four values of $\alpha$ is shown in the top panel of Figure~\ref{fig:sdp_kde}. For $\alpha=0.01$ (dark orange lines) the 10\% and 50\% credible interval contours are invisible in the top-right corner of the panel, as the vast majority of simulated regions have $\rho_+ \approx 1$ and $\mathcal{M} \approx 1$. For $\alpha=0.1$ the KDE spreads out slightly, with the 50\% credible interval reaching $\rho_+\approx0.8$ and $\mathcal{M}\approx0.9$. For $\alpha=1$, the PDF shifts such that the 50\% credible interval contour stretches from $\mathcal{M} = 1$ to $\mathcal{M} \approx 0.8$, while we have $0 \lesssim \rho_+ \lesssim 0.8$. For $\alpha=10$ we find $\rho_+$ centered around zero, while the 50\% credible interval of $\mathcal{M}$ extends to $\mathcal{M} \approx 0.3$. The bottom panel of Figure~\ref{fig:sdp_kde} repeats the exercise for the backward cross-correlation. It confirms that $\rho_-$ and $\mathcal{M}$ are uncorrelated, as predicted elsewhere \citep{Fulgenzi2017, Carlin2019quasi, Carlin2019ac}. Integrating over $\rho_{\pm}$ the reader would find that the marginal distribution of $\mathcal{M}$ is broad in both panels, for $\alpha \gtrsim 1$, as it is difficult to confidently reject the null hypothesis that two sets of samples are from the same distribution, when dealing with small sample sizes. 

Monte Carlo calibration in hand, we present the equivalent data for all regions in the \emph{GOES} catalog with $N_k \geq 10$ in Figure~\ref{fig:goes_kde}. To compute $\mathcal{M}_k$ for each region we first normalize both the waiting times and the sizes by their respective means for that region, before applying the KS two-sample test. In the top (bottom) panel we see no clear relationship between $\rho_{+,\,k}$ ($\rho_{-,\,k}$) and $\mathcal{M}_k$. Marginalizing over $\mathcal{M}_k$ we recover the PDFs of $\rho_{+,\,k}$ and $\rho_{-,\,k}$, shown in the top panel of Figure~\ref{fig:hist_corr}. If, for the sake of argument, we assume that all regions have the same value of $\alpha$, we can compare the KDEs in Figure~\ref{fig:goes_kde} to those in Figure~\ref{fig:sdp_kde}. Under this assumption we are pushed into the regime $1 \lesssim \alpha \lesssim 10$, as the $p(\rho_{+,\,k}, \mathcal{M}_k)$ KDE is slightly off-set from the horizontal axis (i.e.~median $\rho_{+,\,k} > 0$), but the 50\% credible interval for $\mathcal{M}_k$ only extends to $\mathcal{M}_k \lesssim 0.6$. Even if $\alpha$ is not exactly the same in each region, we can interpret the above result as ruling out that a large proportion of regions have $\alpha \lesssim 1$, as if that were the case we would see higher values of $\mathcal{M}_k$ associated with higher $\rho_{+,\,k}$ more often than in Figure~\ref{fig:goes_kde}.

\subsection{Autocorrelations and longer term memory} \label{sec:ac}
While there is a wealth of information available in flare waiting time and size PDFs, as in Section~\ref{sec:disagg}, and cross-correlations, as in Sections~\ref{sec:corr} and \ref{sec:matching}, one may also consider statistics that quantify the longer term memory of the stress in the system. For example, as in the context of neutron star glitches \citep{Carlin2019ac}, studying the autocorrelation between consecutive waiting times, $\rho_{\Delta t}$, or between consecutive sizes, $\rho_{\Delta s}$,  allows one to place constraints on applicable SDP model parameters. In the fast-driven regime, $\alpha \ll 1$, we predict $\rho_{\Delta t} = 0$ and $\rho_{\Delta s} = 0$, as we have $X(t) \rightarrow \xc$ before every stress-release event resulting in the system resetting at every event. If we have $\etadx \propto \delta[X(t_i^-)]$, i.e.~all stress in the system is released at each event, the same prediction of no autocorrelations holds --- again the system is reset at every event. Observations of non-zero autocorrelations therefore rule out certain regimes in the SDP framework.

When we calculate $\rho_{\Delta t,\,k}$ and $\rho_{\Delta s,\,k}$ for all regions in the \emph{GOES} catalog with $N_k \geq 10$, we find that the PDFs $p(\rho_{\Delta t,\,k})$ and $p(\rho_{\Delta s,\,k})$ are broad, akin to $p(\rho_{+,\,k})$ shown in Figure~\ref{fig:hist_corr}. The medians are 0.11 and 0.10 respectively. This result is incongruent with the assumption $\alpha \ll 1$ in all regions, as $\alpha \ll 1$ implies both PDFs should have a median of zero. These results likely stem from the obscuration effects described in Section~\ref{sec:obscuration}. One tentative physical interpretation of a positive $\rho_{\Delta t}$ and $\rho_{\Delta s}$ is via analogy to terrestrial earthquakes which exhibit aftershocks, i.e.~large events are often followed by larger-than-average events, with smaller-than-average waiting times \citep{Utsu1995}. 

\section{Conclusion} \label{sec:conclusion}
In this paper, we revisit the long-standing question of whether solar flares in different active regions are triggered by a stress-relax process with a trigger threshold which is constant in time. We do so by mapping the flaring process to a SDP process, which operates independently in each active region. The SDP framework has been applied in related contexts to forest fires \citep{Daly2006} and solar flares \citep{Wheatland2008}, before being extended in the context of neutron star glitches \citep{Fulgenzi2017, Melatos2018, Carlin2019quasi, Carlin2019ac}. If one assumes a constant driving rate, and a constant ``stress'' threshold at which relaxation events are guaranteed to occur, the model makes precise, falsifiable predictions regarding the statistics of sequences of waiting times and sizes, such as their PDFs, and cross-correlations. It is agnostic about the underlying microphysical mechanism.

Analyzing the historical \emph{GOES} soft X-ray flare catalog, using data from 1975 until 2022, we systematically search all active regions for signatures that flares are consistent with the SDP model with $\alpha \ll 1$ (rapidly driven). We find no evidence that this is the case. Specifically, when considering just the waiting time and size PDFs of each active region we find that either \begin{enumerate*}[i)] \item the conditional PDF of stress-release sizes, $\etadx$, varies from one region to another, or \item a flare rate of $\lambda_k \gtrsim 0.2\,\textrm{hours}^{-1}$ does not correspond to $\alpha \ll 1$ (assuming that $\lambda_k$ traces $\alpha$).\end{enumerate*} The analysis takes into account the selection effect that obscures the detection of flares following a large flare due to enhancement of the soft X-ray background.

On the other hand, if many active regions house a SDP process driven rapidly towards a static-in-time threshold before each event, i.e. have $\alpha \ll 1$, then those regions that have large cross-correlations between event sizes and subsequent waiting times should also have waiting time and size PDFs of the same shape. This prediction is not supported by the data, as proved clearly by Figure~\ref{fig:goes_kde}. The match between $p(\Delta t)$ and $p(\Delta s)$, quantified by $\mathcal{M}_k$ (the $p$-value from a KS two-sample test) for the $k$-th active region, does not correlate with the forward cross-correlation, $\rho_{+,\,k}$. If we assume each region has the same $\alpha$, this test implies $1 \lesssim \alpha \lesssim 10$.

We emphasize that the prediction that $\mathcal{M}_k$ should correlate with $\rho_{+,\,k}$ for a process driven to a threshold does not rely on the specific details of the SDP framework, such as the relationship between flare rate and stress in Equation~\eqref{eq:lam}, nor the functional form of $\etadx$. Any stochastic process that drives the stress towards a static threshold before each event would predict an equivalent observable, for example the Brownian stress-accumulation model \citep{Carlin2020bsa}.

There are many ways to interpret the results in Sections~\ref{sec:agg} and \ref{sec:disagg_general}: \begin{enumerate*}[i)]
	\item solar flares may not be triggered by stress (e.g.~magnetic energy density) breaching a threshold; a completely different process may trigger a flare.
	\item The \emph{GOES} flare catalog is incomplete; flares that occur in the aftermath of large flares are not recorded, creating an artificial cross-correlation.
	\item The threshold at which a glitch is triggered and/or
	\item the driving rate at which stress accumulates may vary with time; either of options iii) and iv) could wash out the observable signature.
\end{enumerate*}
Untangling these explanations entails exploiting the entire wealth of data available in the flare catalog(s), rather than focussing on individual special flares or regions. 

In closing, we touch on the following question: if option iii) above is true, is a threshold that varies with time compatible with any plausible microphysical flare triggers? The question falls outside the scope of this paper, which focuses on the microphysics-agnostic analysis in Sections~\ref{sec:data} and \ref{sec:disagg_general}, so we limit ourselves to the following brief remark. Magnetohydrodynamic instabilities relevant to solar flare activity are triggered above a threshold, which typically depends on the detailed geometry of a flaring magnetic loop, not just its bulk properties (e.g.~magnetic energy density). The geometry of a flaring loop does vary with time in general. For example, the ideal kink instability occurs, when the total twist $\Phi = l B_\phi(r) / [r B_z(r)]$ satisfies $\Phi > \Phi_{\rm cr}(a) \sim 10$, where $l$ and $r$ are the length and minor radius of the current-carrying loop, $B_\phi(r)$ and $B_z(r)$ are the toroidal and axial magnetic field components, and $a$ is the loop aspect ratio \citep{Torok2004}. As sub-photospheric turbulence perturbs randomly the footpoints of a magnetic flux tube, the variables $a$ and hence $\Phi_{\rm cr}(a)$ fluctuate stochastically. Quantifying the amplitude (drift and diffusion) of the fluctuations is a task for detailed magnetohydrodynamic simulations and lies well outside the scope of this paper, but it is conceivable that the amplitude is sufficiently large to render option iii) above viable.

\section*{Acknowledgements}
We acknowledge helpful discussions with Kai Yang in the early stages of this paper. JBC, AM, and MSW are supported by the Australian Research Council (ARC) Centre of Excellence for Gravitational Wave Discovery (OzGrav) (project number CE170100004) and ARC Discovery Project DP220102201. JBC is supported by an Australian Postgraduate Award.

\emph{Software:} \texttt{Numpy} \citep{Harris2020}, \texttt{Scipy} \citep{Virtanen2020}, \texttt{Matplotlib} \citep{Hunter2007}, and \texttt{Stan} \citep{stan2022} through the \texttt{cmdstanpy} interface.

\bibliographystyle{aasjournal}
\bibliography{solar_sdp}

\begin{thebibliography}{}
\expandafter\ifx\csname natexlab\endcsname\relax\def\natexlab#1{#1}\fi
\providecommand{\url}[1]{\href{#1}{#1}}
\providecommand{\dodoi}[1]{doi:~\href{http://doi.org/#1}{\nolinkurl{#1}}}
\providecommand{\doeprint}[1]{\href{http://ascl.net/#1}{\nolinkurl{http://ascl.net/#1}}}
\providecommand{\doarXiv}[1]{\href{https://arxiv.org/abs/#1}{\nolinkurl{https://arxiv.org/abs/#1}}}

\bibitem[{Akaike(1974)}]{Akaike1974}
Akaike, H. 1974, IEEE Transactions on Automatic Control, 19, 716.
\newblock \url{https://ui.adsabs.harvard.edu/abs/1974ITAC...19..716A/abstract}

\bibitem[{Aschwanden \& Johnson(2021)}]{Aschwanden2021a}
Aschwanden, M.~J., \& Johnson, J.~R. 2021, The Astrophysical Journal, 921, 82,
  \dodoi{10.3847/1538-4357/ac2a29}

\bibitem[{Aschwanden {et~al.}(2021)Aschwanden, Johnson, \&
  Nurhan}]{Aschwanden2021}
Aschwanden, M.~J., Johnson, J.~R., \& Nurhan, Y.~I. 2021, The Astrophysical
  Journal, 921, 166, \dodoi{10.3847/1538-4357/ac19a9}

\bibitem[{Aschwanden {et~al.}(2018)Aschwanden, Scholkmann, Béthune, Schmutz,
  Abramenko, Cheung, Müller, Benz, Chernov, Kritsuk, Scargle, Melatos,
  Wagoner, Trimble, \& Green}]{Aschwanden2018}
Aschwanden, M.~J., Scholkmann, F., Béthune, W., {et~al.} 2018, Space Science
  Reviews, 214, 55, \dodoi{10.1007/s11214-018-0489-2}

\bibitem[{Bak {et~al.}(1987)Bak, Tang, \& Wiesenfeld}]{Bak1987}
Bak, P., Tang, C., \& Wiesenfeld, K. 1987, Physical Review Letters, 59, 381,
  \dodoi{10.1103/PhysRevLett.59.381}

\bibitem[{Bak {et~al.}(1988)Bak, Tang, \& Wiesenfeld}]{Bak1988}
---. 1988, Physical Review A, 38, 364, \dodoi{10.1103/PhysRevA.38.364}

\bibitem[{Benz(2016)}]{Benz2016}
Benz, A.~O. 2016, Living Reviews in Solar Physics, 14, 2,
  \dodoi{10.1007/s41116-016-0004-3}

\bibitem[{Betancourt(2018)}]{Betancourt2018}
Betancourt, M. 2018, arXiv e-prints.
\newblock \url{http://arxiv.org/abs/1701.02434}

\bibitem[{Biesecker(1994)}]{Biesecker1994}
Biesecker, D.~A. 1994, PhD thesis, University of New Hampshire.
\newblock \url{https://ui.adsabs.harvard.edu/abs/1994PhDT........51B}

\bibitem[{Boffetta {et~al.}(1999)Boffetta, Carbone, Giuliani, Veltri, \&
  Vulpiani}]{Boffetta1999}
Boffetta, G., Carbone, V., Giuliani, P., Veltri, P., \& Vulpiani, A. 1999,
  Physical Review Letters, 83, 4662, \dodoi{10.1103/PhysRevLett.83.4662}

\bibitem[{Carlin \& Melatos(2019{\natexlab{a}})}]{Carlin2019quasi}
Carlin, J.~B., \& Melatos, A. 2019{\natexlab{a}}, Monthly Notices of the Royal
  Astronomical Society, 483, 4742, \dodoi{10.1093/mnras/sty3433}

\bibitem[{Carlin \& Melatos(2019{\natexlab{b}})}]{Carlin2019ac}
---. 2019{\natexlab{b}}, Monthly Notices of the Royal Astronomical Society,
  488, 4890, \dodoi{10.1093/mnras/stz2014}

\bibitem[{Carlin \& Melatos(2020)}]{Carlin2020bsa}
---. 2020, Monthly Notices of the Royal Astronomical Society, 494, 3383,
  \dodoi{10.1093/mnras/staa935}

\bibitem[{Carlin \& Melatos(2021)}]{Carlin2021endog}
---. 2021, The Astrophysical Journal, 917, 1, \dodoi{10.3847/1538-4357/ac06a2}

\bibitem[{Cox(1955)}]{Cox1955}
Cox, D.~R. 1955, Journal of the Royal Statistical Society, 17, 129,
  \dodoi{10.2307/2983950}

\bibitem[{Crosby {et~al.}(1998)Crosby, Vilmer, Lund, \& Sunyaev}]{Crosby1998}
Crosby, N., Vilmer, N., Lund, N., \& Sunyaev, R. 1998, Astronomy and
  Astrophysics, 334, 299.
\newblock \url{https://ui.adsabs.harvard.edu/abs/1998A&A...334..299C/abstract}

\bibitem[{Daly \& Porporato(2006)}]{Daly2006}
Daly, E., \& Porporato, A. 2006, Physical Review E, 74, 1,
  \dodoi{10.1103/PhysRevE.74.041112}

\bibitem[{Farhang {et~al.}(2019)Farhang, Wheatland, \& Safari}]{Farhang2019}
Farhang, N., Wheatland, M.~S., \& Safari, H. 2019, The Astrophysical Journal,
  883, L20, \dodoi{10.3847/2041-8213/ab40c3}

\bibitem[{Fisher {et~al.}(2012)Fisher, Welsch, \& Abbett}]{Fisher2012}
Fisher, G.~H., Welsch, B.~T., \& Abbett, W.~P. 2012, Solar Physics, 277, 153,
  \dodoi{10.1007/s11207-011-9816-4}

\bibitem[{Fletcher {et~al.}(2011)Fletcher, Dennis, Hudson, Krucker, Phillips,
  Veronig, Battaglia, Bone, Caspi, Chen, Gallagher, Grigis, Ji, Liu, Milligan,
  \& Temmer}]{Fletcher2011}
Fletcher, L., Dennis, B.~R., Hudson, H.~S., {et~al.} 2011, Space Science
  Reviews, 159, 19, \dodoi{10.1007/s11214-010-9701-8}

\bibitem[{Fulgenzi {et~al.}(2017)Fulgenzi, Melatos, \& Hughes}]{Fulgenzi2017}
Fulgenzi, W., Melatos, A., \& Hughes, B.~D. 2017, Monthly Notices of the Royal
  Astronomical Society, 470, 4307, \dodoi{10.1093/mnras/stx1353}

\bibitem[{Gardiner(2009)}]{Gardiner2009}
Gardiner, C. 2009, Stochastic {Methods}: {A} {Handbook} for the {Natural} and
  {Social} {Sciences}, 4th edn. (Berlin Heidelberg: Springer-Verlag).
\newblock \url{https://www.springer.com/gp/book/9783540707127}

\bibitem[{Gavriil {et~al.}(2004)Gavriil, Kaspi, \& Woods}]{Gavriil2004}
Gavriil, F.~P., Kaspi, V.~M., \& Woods, P.~M. 2004, The Astrophysical Journal,
  607, 959, \dodoi{10.1086/383564}

\bibitem[{Gelman {et~al.}(2013)Gelman, Carlin, Stern, Dunson, Vehtari, \&
  Rubin}]{Gelman2013}
Gelman, A., Carlin, J.~B., Stern, H.~S., {et~al.} 2013, Bayesian data analysis,
  3rd edn. (New York: Chapman and Hall/CRC)

\bibitem[{Gorobets \& Messerotti(2012)}]{Gorobets2012}
Gorobets, A., \& Messerotti, M. 2012, Solar Physics, 281, 651,
  \dodoi{10.1007/s11207-012-0121-7}

\bibitem[{Gourdji {et~al.}(2019)Gourdji, Michilli, Spitler, Hessels, Seymour,
  Cordes, \& Chatterjee}]{Gourdji2019}
Gourdji, K., Michilli, D., Spitler, L.~G., {et~al.} 2019, The Astrophysical
  Journal Letters, 877, L19, \dodoi{10.3847/2041-8213/ab1f8a}

\bibitem[{Göğüş {et~al.}(1999)Göğüş, Woods, Kouveliotou, Paradijs,
  Briggs, Duncan, \& Thompson}]{Gogus1999}
Göğüş, E., Woods, P.~M., Kouveliotou, C., {et~al.} 1999, The Astrophysical
  Journal, 526, L93, \dodoi{10.1086/312380}

\bibitem[{Göğüş {et~al.}(2000)Göğüş, Woods, Kouveliotou, Paradijs,
  Briggs, Duncan, \& Thompson}]{Gogus2000}
---. 2000, The Astrophysical Journal, 532, L121, \dodoi{10.1086/312583}

\bibitem[{Harris {et~al.}(2020)Harris, Millman, van~der Walt, Gommers,
  Virtanen, Cournapeau, Wieser, Taylor, Berg, Smith, Kern, Picus, Hoyer, van
  Kerkwijk, Brett, Haldane, del Río, Wiebe, Peterson, Gérard-Marchant,
  Sheppard, Reddy, Weckesser, Abbasi, Gohlke, \& Oliphant}]{Harris2020}
Harris, C.~R., Millman, K.~J., van~der Walt, S.~J., {et~al.} 2020, Nature, 585,
  357, \dodoi{10.1038/s41586-020-2649-2}

\bibitem[{Haskell \& Melatos(2015)}]{Haskell2015}
Haskell, B., \& Melatos, A. 2015, International Journal of Modern Physics D,
  24, 1530008, \dodoi{10.1142/S0218271815300086}

\bibitem[{Hudson(2019)}]{Hudson2019}
Hudson, H.~S. 2019, Monthly Notices of the Royal Astronomical Society, stz3121,
  \dodoi{10.1093/mnras/stz3121}

\bibitem[{Hudson(2020)}]{Hudson2020}
---. 2020, Solar Physics, 295, 132, \dodoi{10.1007/s11207-020-01698-w}

\bibitem[{Hudson {et~al.}(1998)Hudson, Labonte, Sterling, \&
  Watanabe}]{Hudson1998}
Hudson, H.~S., Labonte, B.~J., Sterling, A.~C., \& Watanabe, T. 1998, in Space
  {Science} {Library}, Vol. 229, Observational {Plasma} {Astrophysics}: {Five}
  {Years} of {Yohkoh} and {Beyond} (Springer Dordrecht), 237.
\newblock \url{https://doi.org/10.1007/978-94-011-5220-4}

\bibitem[{Hunter(2007)}]{Hunter2007}
Hunter, J.~D. 2007, Computing in Science and Engineering, 9, 90,
  \dodoi{10.1109/MCSE.2007.55}

\bibitem[{Hurvich \& Tsai(1989)}]{Hurvich1989}
Hurvich, C.~M., \& Tsai, C.-L. 1989, Biometrika, 76, 297,
  \dodoi{10.1093/biomet/76.2.297}

\bibitem[{Jensen(1998)}]{Jensen1998}
Jensen, H.~J. 1998, Self-{Organized} {Criticality}. {Emergent} {Complex}
  {Behavior} in {Physical} and {Biological} {Systems}, Cambridge {Lecture}
  {Notes} in {Physics} (Cambridge: Cambridge University Press)

\bibitem[{Ji \& Daughton(2011)}]{Ji2011}
Ji, H., \& Daughton, W. 2011, Physics of Plasmas, 18, 111207,
  \dodoi{10.1063/1.3647505}

\bibitem[{Kanazir \& Wheatland(2010)}]{Kanazir2010}
Kanazir, M., \& Wheatland, M.~S. 2010, Solar Physics, 266, 301,
  \dodoi{10.1007/s11207-010-9623-3}

\bibitem[{Keys {et~al.}(2011)Keys, Mathioudakis, Jess, Shelyag, Crockett,
  Christian, \& Keenan}]{Keys2011}
Keys, P.~H., Mathioudakis, M., Jess, D.~B., {et~al.} 2011, The Astrophysical
  Journal Letters, 740, L40, \dodoi{10.1088/2041-8205/740/2/L40}

\bibitem[{Kingman(1993)}]{Kingman1993}
Kingman, J. F.~C. 1993, Poisson processes (Oxford: Oxford University Press)

\bibitem[{La~Roche-Carrier {et~al.}(2019)La~Roche-Carrier, Dituba~Ngoma,
  Kocaefe, \& Erchiqui}]{LaRoche-Carrier2019}
La~Roche-Carrier, N., Dituba~Ngoma, G., Kocaefe, Y., \& Erchiqui, F. 2019,
  International Journal of Quality \& Reliability Management, 37, 223,
  \dodoi{10.1108/IJQRM-01-2019-0035}

\bibitem[{Last \& Penrose(2017)}]{Last2017}
Last, G., \& Penrose, M. 2017, Lectures on the {Poisson} {Process} (Cambridge
  University Press), \dodoi{10.1017/9781316104477}

\bibitem[{Lehmann \& D'Abrera(2006)}]{Lehmann2006}
Lehmann, E.~L., \& D'Abrera, H. J.~M. 2006, Nonparametrics: {Statistical}
  {Methods} {Based} on {Ranks} (New York: Springer-Verlag)

\bibitem[{Lepreti {et~al.}(2001)Lepreti, Carbone, \& Veltri}]{Lepreti2001}
Lepreti, F., Carbone, V., \& Veltri, P. 2001, The Astrophysical Journal, 555,
  L133, \dodoi{10.1086/323178}

\bibitem[{Li \& Fenimore(1996)}]{Li1996}
Li, H., \& Fenimore, E.~E. 1996, The Astrophysical Journal, 469, L115,
  \dodoi{10.1086/310275}

\bibitem[{Lippiello {et~al.}(2010)Lippiello, Arcangelis, \&
  Godano}]{Lippiello2010}
Lippiello, E., Arcangelis, L.~d., \& Godano, C. 2010, Astronomy and
  Astrophysics, 511, L2, \dodoi{10.1051/0004-6361/200913784}

\bibitem[{Lu(1995{\natexlab{a}})}]{Lu1995}
Lu, E.~T. 1995{\natexlab{a}}, The Astrophysical Journal, 446, L109,
  \dodoi{10.1086/187942}

\bibitem[{Lu(1995{\natexlab{b}})}]{Lu1995a}
---. 1995{\natexlab{b}}, The Astrophysical Journal, 447, 416,
  \dodoi{10.1086/175885}

\bibitem[{Lu \& Hamilton(1991)}]{Lu1991}
Lu, E.~T., \& Hamilton, R.~J. 1991, The Astrophysical Journal Letters, 380,
  L89, \dodoi{10.1086/186180}

\bibitem[{Lu {et~al.}(1993)Lu, Hamilton, McTiernan, \& Bromund}]{Lu1993}
Lu, E.~T., Hamilton, R.~J., McTiernan, J.~M., \& Bromund, K.~R. 1993, The
  Astrophysical Journal, 412, 841, \dodoi{10.1086/172966}

\bibitem[{Lyne \& Graham-Smith(2012)}]{Lyne2012}
Lyne, A., \& Graham-Smith, F. 2012, Pulsar {Astronomy}, 4th edn., Cambridge
  {Astrophysics} (Cambridge: Cambridge University Press),
  \dodoi{10.1017/CBO9780511844584}

\bibitem[{Melatos \& Drummond(2019)}]{Melatos2019}
Melatos, A., \& Drummond, L.~V. 2019, The Astrophysical Journal, 885, 37,
  \dodoi{10.3847/1538-4357/ab44c3}

\bibitem[{Melatos {et~al.}(2018)Melatos, Howitt, \& Fulgenzi}]{Melatos2018}
Melatos, A., Howitt, G., \& Fulgenzi, W. 2018, The Astrophysical Journal, 863,
  196, \dodoi{10.3847/1538-4357/aad228}

\bibitem[{Millhouse {et~al.}(2022)Millhouse, Melatos, Howitt, Carlin, Dunn, \&
  Ashton}]{Millhouse2022}
Millhouse, M., Melatos, A., Howitt, G., {et~al.} 2022, Monthly Notices of the
  Royal Astronomical Society, 511, 3304, \dodoi{10.1093/mnras/stac194}

\bibitem[{Mitzenmacher(2003)}]{Mitzenmacher2003}
Mitzenmacher, M. 2003, Internet Mathematics, 1, 226.
\newblock
  \url{https://projecteuclid.org/journals/internet-mathematics/volume-1/issue-2/A-Brief-History-of-Generative-Models-for-Power-Law-and/im/1089229510.full}

\bibitem[{Paxson \& Floyd(1994)}]{Paxson1994}
Paxson, V., \& Floyd, S. 1994, ACM SIGCOMM Computer Communication Review, 24,
  257, \dodoi{10.1145/190809.190338}

\bibitem[{Peng \& Zhao(2009)}]{Peng2009}
Peng, Z., \& Zhao, P. 2009, Nature Geoscience, 2, 877, \dodoi{10.1038/ngeo697}

\bibitem[{Priest \& Forbes(2002)}]{Priest2002}
Priest, E., \& Forbes, T. 2002, The Astronomy and Astrophysics Review, 10, 313,
  \dodoi{10.1007/s001590100013}

\bibitem[{Rosner \& Vaiana(1978)}]{Rosner1978}
Rosner, R., \& Vaiana, G.~S. 1978, The Astrophysical Journal, 222, 1104,
  \dodoi{10.1086/156227}

\bibitem[{Sahu {et~al.}(2022)Sahu, Joshi, Prasad, \& Cho}]{Sahu2022}
Sahu, S., Joshi, B., Prasad, A., \& Cho, K.-S. 2022, Evolution of magnetic
  fields and energy release processes during homologous eruptive flares,
  arXiv, \dodoi{10.48550/arXiv.2212.04150}

\bibitem[{Shelyag {et~al.}(2011{\natexlab{a}})Shelyag, Fedun, Keenan, Erdélyi,
  \& Mathioudakis}]{Shelyag2011a}
Shelyag, S., Fedun, V., Keenan, F.~P., Erdélyi, R., \& Mathioudakis, M.
  2011{\natexlab{a}}, Annales Geophysicae, 29, 883,
  \dodoi{10.5194/angeo-29-883-2011}

\bibitem[{Shelyag {et~al.}(2011{\natexlab{b}})Shelyag, Keys, Mathioudakis, \&
  Keenan}]{Shelyag2011}
Shelyag, S., Keys, P., Mathioudakis, M., \& Keenan, F.~P. 2011{\natexlab{b}},
  Astronomy and Astrophysics, 526, A5, \dodoi{10.1051/0004-6361/201015645}

\bibitem[{Shibata \& Tanuma(2001)}]{Shibata2001}
Shibata, K., \& Tanuma, S. 2001, Earth, Planets and Space, 53, 473,
  \dodoi{10.1186/BF03353258}

\bibitem[{Singhai {et~al.}(2007)Singhai, Joshi, \& Bhatt}]{Singhai2007}
Singhai, R., Joshi, S.~D., \& Bhatt, R. K.~P. 2007, in 2007 15th
  {International} {Conference} on {Software}, {Telecommunications} and
  {Computer} {Networks}, 1--5, \dodoi{10.1109/SOFTCOM.2007.4446103}

\bibitem[{{Stan Development Team}(2022)}]{stan2022}
{Stan Development Team}. 2022, Stan {Modeling} {Language} {Users} {Guide} and
  {Reference} {Manual}, 2.30.
\newblock \url{https://mc-stan.org}

\bibitem[{Sun {et~al.}(2017)Sun, Hoeksema, Liu, Kazachenko, \& Chen}]{Sun2017}
Sun, X., Hoeksema, J.~T., Liu, Y., Kazachenko, M., \& Chen, R. 2017, The
  Astrophysical Journal, 839, 67, \dodoi{10.3847/1538-4357/aa69c1}

\bibitem[{Sun {et~al.}(2012)Sun, Hoeksema, Liu, Wiegelmann, Hayashi, Chen, \&
  Thalmann}]{Sun2012}
Sun, X., Hoeksema, J.~T., Liu, Y., {et~al.} 2012, The Astrophysical Journal,
  748, 77, \dodoi{10.1088/0004-637X/748/2/77}

\bibitem[{Török {et~al.}(2004)Török, Kliem, \& Titov}]{Torok2004}
Török, T., Kliem, B., \& Titov, V.~S. 2004, Astronomy and Astrophysics, 413,
  L27, \dodoi{10.1051/0004-6361:20031691}

\bibitem[{Utsu {et~al.}(1995)Utsu, Ogata, S, \& {Matsu'ura}}]{Utsu1995}
Utsu, T., Ogata, Y., S, R., \& {Matsu'ura}. 1995, Journal of Physics of the
  Earth, 43, 1, \dodoi{10.4294/jpe1952.43.1}

\bibitem[{van~der Linden(2006)}]{VanDerLinden2006}
van~der Linden, W.~J. 2006, Journal of Educational and Behavioral Statistics,
  31, 181.
\newblock \url{https://www.jstor.org/stable/3701364}

\bibitem[{Verbeeck {et~al.}(2019)Verbeeck, Kraaikamp, Ryan, \&
  Podladchikova}]{Verbeeck2019}
Verbeeck, C., Kraaikamp, E., Ryan, D.~F., \& Podladchikova, O. 2019, The
  Astrophysical Journal, 884, 50, \dodoi{10.3847/1538-4357/ab3425}

\bibitem[{Virtanen {et~al.}(2020)Virtanen, Gommers, Oliphant, Haberland, Reddy,
  Cournapeau, Burovski, Peterson, Weckesser, Bright, van~der Walt, Brett,
  Wilson, Millman, Mayorov, Nelson, Jones, Kern, Larson, Carey, Polat, Feng,
  Moore, VanderPlas, Laxalde, Perktold, Cimrman, Henriksen, Quintero, Harris,
  Archibald, Ribeiro, Pedregosa, \& van Mulbregt}]{Virtanen2020}
Virtanen, P., Gommers, R., Oliphant, T.~E., {et~al.} 2020, Nature Methods, 17,
  261, \dodoi{10.1038/s41592-019-0686-2}

\bibitem[{Wand \& Jones(1995)}]{Wand1995}
Wand, M.~P., \& Jones, M.~C. 1995, Kernel {Smoothing} (London; New York:
  Chapman \& Hall)

\bibitem[{Wheatland(2000{\natexlab{a}})}]{Wheatland2000corr}
Wheatland, M.~S. 2000{\natexlab{a}}, Solar Physics, 191, 381,
  \dodoi{10.1023/A:1005240712931}

\bibitem[{Wheatland(2000{\natexlab{b}})}]{Wheatland2000delt}
---. 2000{\natexlab{b}}, The Astrophysical Journal, 536, L109,
  \dodoi{10.1086/312739}

\bibitem[{Wheatland(2001)}]{Wheatland2001}
---. 2001, Solar Physics, 203, 87, \dodoi{10.1023/A:1012749706764}

\bibitem[{Wheatland(2008)}]{Wheatland2008}
---. 2008, The Astrophysical Journal, 679, 1621, \dodoi{10.1086/587871}

\bibitem[{Wheatland \& Glukhov(1998)}]{Wheatland1998}
Wheatland, M.~S., \& Glukhov, S. 1998, The Astrophysical Journal, 494, 858,
  \dodoi{10.1086/305245}

\end{thebibliography}

\appendix
\section{Magnetic energy density as a stress variable in an active region} \label{app:toy}
The SDP framework described in Sections~\ref{sec:sdp} and \ref{sec:pred} is agnostic about the microphysics of solar flares, beyond the general assumption that the flaring rate increases with the stress in the system and diverges at a stress threshold. In this appendix, we sketch briefly one possible mapping between the SDP model and a specific microphysical flare model, in which the stress variable is the spatially-averaged magnetic energy density in an active region. We emphasize that the model is phenomenological and highly idealized. We do not favor it over the many alternatives; it is merely one possible illustration of how such a mapping may work, as a guide to the interested reader.

Let $X(t)$ correspond to the spatially-averaged magnetic energy density in an active region. The spatial average is taken in order to package the stress into a single variable, noting that the magnetic energy density is nonuniform in reality. Let the active region have characteristic linear dimension $L$. Let $S$ be the magnetohydrodynamic Poynting flux through the photosphere. We assume that the Poynting flux deposits magnetic energy into the active region without losses and at a constant rate, so that one has $X(t) = X(t_i^+) + S t /L$ between flares ($t_i^+ \leq t \leq t_{i+1}^-$). Recent vector magnetogram measurements allow direct observation of the Poynting flux, and the magnetic energy density of an active region, which are at-odds with our assumption that $S$ is steady in time \citep{Fisher2012, Sun2012, Sun2017, Sahu2022}. Putting aside the latter consideration for the moment, we can write the control parameter $\alpha$ as
\begin{equation}
\alpha = \frac{\lambda_0 S}{\xc L}\ ,
\end{equation}
where $X_{\rm cr}$ is the magnetic energy density threshold for a magnetohydrodynamic instability, for example, and $\lambda_0$ is the instability's trigger reference rate.

Several plausible magnetohydrodynamic instabilities have been suggested as solar flare triggers in the literature \citep{Ji2011}. Some of them do not involve a magnetic energy density threshold at all. For example, the kink instability discussed at the end of Section~\ref{sec:conclusion} is triggered when the field-line twist $\Phi$ exceeds a threshold $\Phi_{\rm cr}(a)$, whose value depends on the aspect ratio $a$ of the flaring loop \citep{Torok2004}. However, instabilities triggered by a magnetic energy density threshold do exist. One example is plasmoid-induced magnetic reconnection via tearing modes in a fractal current sheet \citep{Shibata2001}, whose threshold depends on fractional powers of the Alfv\'{e}n speed (or equivalently the Lundquist number) and hence on the magnetic energy density; see \citet{Ji2011} or section 5 in \citet{Shibata2001} for example. In the latter reference, the threshold condition also depends on the aspect ratio $a$ (width divided by length) of the current sheet, which can vary with time, as a flaring loop responds to sub-photospheric turbulence.

What are the time-scales on which $S$ and $\xc$ vary? In the SDP picture, both variables are steady, but the \emph{GOES} analysis in Section~\ref{sec:disagg_general} implies that one or both may vary in reality (although other scenarios are possible too, as discussed in Section~\ref{sec:conclusion}). As far as $S$ is concerned, one expects to find statistical fluctuations on the eddy turnover time-scale of sub-photospheric turbulence and flux emergence, $\tau_{\rm flux}$, as observed with vector magnetograms and Doppler measurements \citep{Fisher2012}. Magnetohydrodynamic simulations and G-band radiative signatures suggest $\tau_{\rm flux} \sim {\rm minutes}$ for magnetic features associated with solar granulation, with peak-to-peak fluctuation amplitude $\lesssim 30\%$ \citep{Keys2011, Shelyag2011, Shelyag2011a}. As far as $\xc$ is concerned, in the plasmoid-induced reconnection picture as one illustrative example, the magnetic energy density threshold for a Sweet-Parker current sheet to undergo secondary tearing depends not only on the magnetic diffusivity, which fluctuates on the same time-scale as the local temperature, but also on $a$ and $L$, which fluctuate on the turnover time-scale $\tau_{\rm flux}$, e.g.~equation (15) in \citep{Shibata2001}. Flare waiting times are typically comparable to or longer than $\tau_{\rm flux}$, so it is conceivable that the constant-$\xc$ approximation in the SDP theory does not apply to every active region.

We emphasize again that the mapping in this appendix is idealized and illustrative. None of the statistical analysis in Sections~\ref{sec:agg} and \ref{sec:disagg_general} is predicated on the microphysics in this appendix.

\newpage
\section{Hierarchical Bayesian framework} \label{app:hierarch}
This appendix lays out a complementary approach to the heuristic tests with which we analyze individual active regions in Section~\ref{sec:disagg_general}. We first write down the likelihood for a set of observed waiting times and sizes in an individual region, given a set of model parameters, in Appendix~\ref{app:likelihood}. We introduce the hierarchical Bayesian framework which combines inference in different regions to estimate population-level parameters, in Appendix~\ref{app:sub_hier}. In Appendix~\ref{app:fake} we test the efficacy of this approach with synthetic data, and explain why it is not appropriate in this paper to apply this framework to the real \emph{GOES} catalog despite its efficacy. Finally, in Appendix~\ref{app:avet} we propose an alternative approach using only the average waiting time in each region, as a motivation for future studies. The recipes in this appendix are included for completeness and as a starting point for readers, who wish to develop hierarchical methods of solar flare analysis further. 

\subsection{Likelihood and Bayes' theorem} \label{app:likelihood}
Solving Equations \eqref{eq:xoft}--\eqref{eq:pdelt} for the long-term observable PDFs, $p(\Delta t)$ and $p(\Delta X)$, is intractable analytically for most choices of $\etadx$. However, for the special case
\begin{equation}
\etadx \propto \left[X(t_i^-) - \Delta X_i \right]^\delta\,, \label{eq:sep_eta}
\end{equation}
we obtain the analytic result
\begin{equation}
p(z) = (\alpha + \delta + 1) (1 - z)^{\alpha + \delta},\, \label{eq:sep_pdfs}
\end{equation}
where the variable $z$ is either $\Delta t$ or $\Delta X$, i.e.~the waiting time and size PDFs are identical \citep{Fulgenzi2017}. In Equation~\eqref{eq:sep_eta} we have $\delta > 0$ and the constant of proportionality is set by the condition that $\eta$ must integrate to unity between $\Delta X_i = 0$ and $\Delta X_i = X(t_i^-)$; see section 6 and appendix D of \citet{Fulgenzi2017} for a full derivation. Equation~\eqref{eq:sep_eta} is a monotonically decreasing function of $\Delta X_i$, i.e.~small stress-release events are preferred over large events. Its specific functional form is reasonable but arbitrary; it is not inferred from solar flare data. We adopt it here as a pedagogical device to illustrate the advantages and disadvantages of a hierarchical Bayesian approach to analyzing flare data.

When we restore the dimensions to Equation~\eqref{eq:sep_pdfs}, and consider the set of observations in the $k$-th active region, $D_k = \{\Delta t_{i,\,k}, \Delta s_{i,\,k}\}$, with $1 \leq i \leq N_k$, we can write the likelihood
\begin{align}
\mathcal{L}(D_k\,|\,\boldsymbol{\theta}_k) ={}& \prod_{i=1}^{N_k - 1} p(\Delta t_{i,\,k}\,|\,\boldsymbol{\theta}_k)\, \prod_{i=1}^{N_k} p(\Delta s_{i,\,k}\,|\,\boldsymbol{\theta}_k) \\
={}& \tau_k^{1 - N_k}\, \xi_k^{-N_k} \left(\beta_k + 1 \right)^{2N_k - 1} \prod_{i=1}^{N_k - 1} \left(1 - \frac{\Delta t_{i,\,k}}{\tau_k}\right)^{\beta_k} \prod_{i=1}^{N_k}\left(1 - \frac{\Delta s_{i,\,k}}{\xi_k} \right)^{\beta_k} \label{eq:full_l}
\end{align}
where $\xi_k$ is the constant of proportionality in Equation~\eqref{eq:delx_prop} which translates the drop in magnetic energy density $\Delta X$ to the observed flare size, $\Delta s$, and one has $\beta_k = \lambda_{0,\,k}\, \tau_k + \delta_k$. While there are $N_k$ flare sizes in the region, there are only $N_k - 1$ waiting times. The three model parameters $\boldsymbol{\theta}_k = \{\tau_k,\, \xi_k,\, \beta_k \}$ are assumed constant in time for the lifetime of the active region. Bayes' theorem calculates the posterior probability distribution (henceforth ``posterior''), $p(\boldsymbol{\theta}_k\,|\,D_k)$, i.e.~the probability density of parameter vector $\boldsymbol{\theta}_k$ given the data $D_k$, using the prior probability distribution (henceforth ``prior''),
 $\pi(\boldsymbol{\theta}_k)$, viz.
\begin{equation}
p(\boldsymbol{\theta}_k\,|\,D_k) \propto \mathcal{L}(D_k\,|\,\boldsymbol{\theta}_k) \pi(\boldsymbol{\theta}_k)\ .
\end{equation}
The normalizing constant of proportionality is often called the evidence and, while essential for model comparison, is not relevant for the parameter estimation exercise below \citep{Gelman2013}.

\subsection{Population-level parameter estimation} \label{app:sub_hier}
Suppose, for the sake of illustration, that $\xc$, $\lambda_0$, and every component of $\boldsymbol{\theta}_k$ are approximately the same in all active regions. In the language of hierarchical Bayesian inference, this corresponds to assuming that the value of $\boldsymbol{\theta}_k$ for each region is a random number drawn from a population-level distribution known as a ``hyper-prior'', with narrow extent. For concreteness, we assume the hyper-prior for each model parameter is a Gaussian with mean and standard deviation $\mu_a$ and $\sigma_a$ respectively, with $a \in \{\tau, \xi, \beta \}$. While $\beta_k$ does depend on $\tau_k$, the inference remains accurate so long as the covariance between $\tau_k$, $\lambda_{0,\,k}$, and $\delta_k$ is minimal. The marginal posterior distribution for the parameters describing the hyper-priors is calculated as
\begin{equation}
p(\boldsymbol{\Lambda}\,|\, \mathcal{D}) \propto \pi(\boldsymbol{\Lambda}) \prod_k^M \int \textrm{d}\boldsymbol{\theta}_k\, \mathcal{L}(D_k\,|\,\boldsymbol{\theta}_k) \pi( \boldsymbol{\theta}_k\,|\,\boldsymbol{\Lambda})\,, \label{eq:marg_post}
\end{equation}
with $\boldsymbol{\Lambda} = \{\mu_{a},\,\sigma_{a} \}$. In Equation~\eqref{eq:marg_post}, $\pi(\boldsymbol{\Lambda})$ is the prior on $\boldsymbol{\Lambda}$, and $\mathcal{D} = \{D_1,...,D_k,...D_M \}$ is the set of all data, $D_k$, from $M$ different active regions.

In practice, we can use a Monte Carlo sampler to estimate Equation~\eqref{eq:marg_post}, and thus infer the posterior predictive distributions, $p(\boldsymbol{\theta})$, where we drop the subscript $k$ to signify that these distributions bring together information from all active regions to inform the inference. We opt to sample using a Hamiltonian Monte Carlo No U-Turn Sampler \citep{Betancourt2018}, as implemented in the \texttt{Stan} programming language \citep{stan2022}. 

\subsection{Validation with synthetic data} \label{app:fake}
To estimate the efficacy of the scheme in Appendix~\ref{app:sub_hier} to infer population-level parameters, we first apply it to a set of synthetic data generated directly from the model. That is, we generate $M=50$ fake ``active regions'', each with $N_k = 100$ events, with flare sizes and waiting times generated from the SDP framework with $\etadx$ as in Equation~\eqref{eq:sep_eta}. In what follows we use the compact notation $x \sim \mathcal{N}(\mu,\, \sigma)$ to denote that the random variable $x$ is drawn from a Gaussian distribution with mean $\mu$ and standard deviation $\sigma$. To generate the data for each region we select a value $\tau_k \sim \mathcal{N}(\mu_\tau = 3\,\textrm{day},\, \sigma_\tau = 0.5\,\textrm{day})$, a value $\xi_k \sim \mathcal{N}(\mu_\xi = 2\,\textrm{arb. units},\, \sigma_\xi = 0.3\,\textrm{arb. units})$, and a value $\beta_k \sim \mathcal{N}(\mu_\beta = 5,\, \sigma_\beta = 0.3)$. Each Gaussian is truncated such that all variates are positive. 

\begin{figure}
\centering
\includegraphics[width=0.8\linewidth]{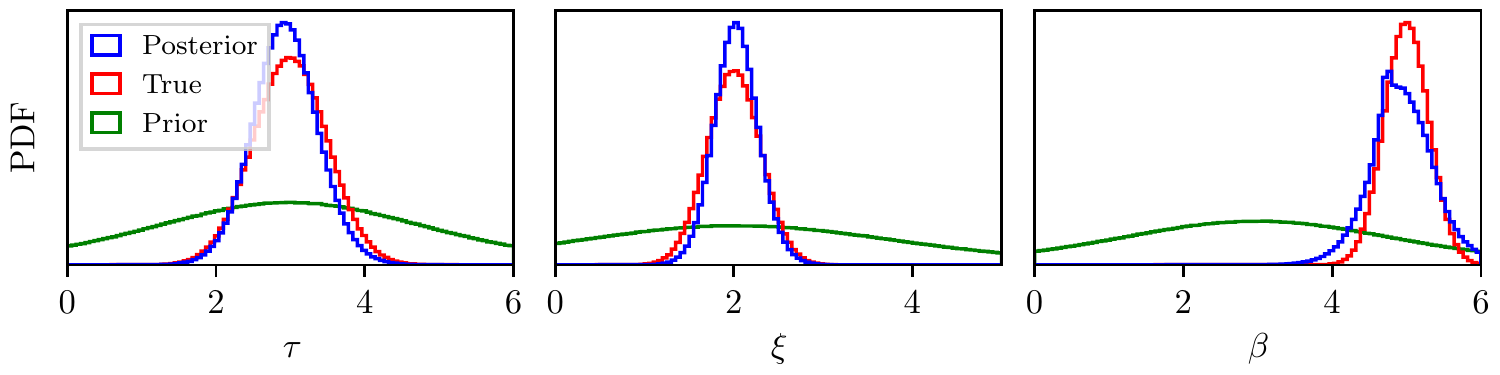}
\caption{Posterior predictive check demonstrating the recovery of injected population-level parameters. The weakly informative priors set on the hyper-parameters $\tau$, $\xi$, and $\beta$ are shown in green, histograms of samples from the injected population parameters are shown in red, while histograms of samples from the posterior distributions are shown in blue. See text in Appendix~\ref{app:fake} for details on the procedure.}
\label{fig:ppc_fake}
\end{figure}

After running the sampler with the synthetic data we perform a posterior predictive check, i.e.~compare the samples from our posterior distributions $p(\tau)$, $p(\xi)$, and $p(\beta)$ with the injected distributions and the priors on the hyper-parameters. We show the results in Figure~\ref{fig:ppc_fake}. We see that the posterior samples (blue) appropriately overlap with the injected ground truth (red) for $\tau$, $\xi$, and $\beta$.  

Despite the encouraging results in Figure~\ref{fig:ppc_fake}, we do not apply this hierarchical Bayesian scheme to the \emph{GOES} catalog in this paper. This is because in most regions neither the waiting time nor the size PDFs follow the functional form in Equation~\eqref{eq:sep_pdfs}. When we include Equation~\eqref{eq:sep_pdfs} in the set of options available for the AICc to select between, as in Section~\ref{sec:disagg}, fewer than $0.1\%$ of active regions are fitted best with Equation~\eqref{eq:sep_pdfs}. Therefore Equation~\eqref{eq:full_l} is not an appropriate likelihood for the data. Other choices of the jump distribution in Equation~\eqref{eq:sep_eta} may alleviate this problem but they lie outside the scope of this paper.

\subsection{Likelihood based on the average waiting time} \label{app:avet}
An alternative approach is to build the likelihood out of a summary statistic, such as the average waiting time in a given region, $\langle \Delta t \rangle_k$, rather than each observed waiting time and/or size. In the SDP framework, we find  \citep{Fulgenzi2017, Millhouse2022}
\begin{equation}
\langle \Delta t \rangle^{\rm SDP} = \frac{1}{\alpha + \alpha_0}\ , \label{eq:sdp_avet}
\end{equation}
where $\alpha_0 \sim 1$ is a dimensionless constant, whose exact value depends on the particular form of $\etadx$. Restoring the dimensions to Equation~\eqref{eq:sdp_avet} we can write the likelihood as
\begin{equation}
\mathcal{L}\left(\langle \Delta t \rangle_k\,|\, \boldsymbol{\theta}_k\right) = \frac{1}{\sqrt{2\pi\sigma_k}} \exp\left[\frac{-\left(\langle \Delta t \rangle_k - \frac{\tau_k}{\lambda_{0,\,k} \tau_k + \alpha_{0,\,k}}\right)^2}{2\sigma_k^2} \right]\ , \label{eq:avet_l}
\end{equation}
with $\boldsymbol{\theta}_k = \{\tau_k,\, \lambda_{0,\,k},\, \alpha_{0,\,k},\, \sigma_k\}$, where we assume the residual of $\langle \Delta t \rangle_k$ with the model subtracted is normally distributed, with zero mean and standard deviation $\sigma_k$. With the likelihood in Equation~\eqref{eq:avet_l} one could perform population-level parameter estimation, as described in Section~\ref{app:sub_hier}. We leave this, and an exploration of how $\boldsymbol{\theta}_k$ depends on parameters intrinsic to each active region, to future work.

\section{Cleaning \emph{GOES} flare summary data} \label{app:cleaning}
The SWPC and NGDC File Transfer Protocol servers that host the \emph{GOES} flare summary data, as well as other data portals such as the Heliophysics Event Knowledgebase\footnote{Accessible through \url{http://hec.helio-vo.eu/hec/hec_gui.php}.}, contain numerous typographical anomalies found when collating the data into a homogeneous catalog. These anomalies are found in the recorded active region number by checking whether one has $\Delta t_{i,\,k} < 14\,\textrm{days}$ for $1 \leq i \leq N_k - 1$ for each active region. The limit of 14 days applies because active regions are typically only visible for two weeks due to the rotation period of the Sun. The one exception in the \emph{GOES} flare summary data is eight flares that occurred on 2002-11-02, and are assigned to active region number 10198. Two weeks later the active region appears on the eastern limb of the Sun, and 33 additional flares are assigned to active region number 10198, from 2002-11-17 until 2002-11-28. We opt not to consider the former eight flares as part of active region number 10198, and do not include them in the catalog.

The anomalies, and hand-corrected values, are tabulated in Table~\ref{tab:cleaning}. The corrected values are determined manually by referring to the context of surrounding flares in the database. For example, the flare starting at 1981-07-18 11:46 is recorded with active region number 3121. Yet the other 11 flares associated with the latter region occurred between 1981-05-24 and 1981-06-01, while active region number 3221 has 32 flares recorded between 1981-07-17 and 1981-07-29, indicating that the anomalous flare should be associated with the latter active region. The latitude of the anomalous flare is also within $\pm 5^{\circ}$ of other flares from active region number 3221, while flares in active region number 3121 are $\gtrsim 10^{\circ}$ higher in latitude. 

\begin{table}
\centering
\caption{Anomalous active region numbers found in the \emph{GOES} flare summary data by checking whether one has $\Delta t_{i,\,k} < 14\,\textrm{days}$ for $1 \leq i \leq N_k - 1$ for each active region. Corrected values are determined manually, by considering the context of what active regions are present at the time of the anomalous flare at a similar latitude. Corrected values of ``---'' indicate that we cannot identify a reasonable active region to associate with the anomalous flare. } \label{tab:cleaning}
{\renewcommand{\arraystretch}{1.3}% for the vertical padding
\begin{ruledtabular}
\begin{tabular}{l l l}
Flare start time & Anomalous active region number & Corrected active region number \\
\midrule
1978-05-30 06:19 & 1000  & 1134  \\
1981-07-18 11:46 & 3121  & 3221  \\
1981-08-22 06:58 & 366   & 3266  \\
1983-03-01 18:24 & 2102  & 4102  \\
1983-03-01 18:54 & 2102  & 4102  \\
1983-07-04 06:09 & 4135  & 4235  \\
1983-07-29 03:53 & 4236  & 4263  \\
1993-09-27 01:35 & 7500  & 7590  \\
2000-11-09 21:13 & 9125  & ---   \\
2002-06-14 20:18 & 1     & 10001 \\
2003-07-31 07:59 & 422   & 10422 \\
2003-12-21 04:10 & 10000 & ---   \\
2011-12-22 13:04 & 11281 & 11381 \\
2017-07-11 01:09 & 12655 & 12665 \\
2017-07-16 10:25 & 12655 & 12665 \\
2021-05-10 23:46 & 12282 & 12822 \\
2021-09-01 03:03 & 12680 & 12860 \\
2021-09-01 04:27 & 12680 & 12860 \\
2021-11-06 22:01 & 12984 & 12894 \\
2021-12-18 11:17 & 12807 & 12907 \\
2021-12-18 17:27 & 12807 & 12907 \\
\end{tabular}
\end{ruledtabular}
}
\end{table}

\section{Log-normal waiting time distributions} \label{app:ln}
A sequence of instantaneous events ordered in time is known as a point process. If the instantaneous event rate does not change with time, and if the process is memoryless, it is called a Poisson point process, and exhibits exponentially distributed waiting times \citep{Kingman1993}. If the instantaneous rate does change with time it is a non-homogeneous Poisson process, which can exhibit waiting times that are distributed as a log-normal for certain rate functions \citep{Gardiner2009,Last2017}. The SDP is one example of a non-homogeneous Poisson process that can generate waiting times distributed as a log-normal, for certain choices of $\etadx$ and $\alpha$. One can alternatively generate log-normally distributed values via a multiplicative process, e.g.~an organism that grows in proportion to its current size multiplied by a random variable generates (at least approximately) log-normally distributed sizes, due to the Central Limit Theorem \citep{Mitzenmacher2003}. 

Log-normal waiting time distributions are observed in many other contexts besides solar flares, including but not limited to gamma-ray bursts \citep{Li1996}, X-ray bursts \citep{Gogus1999,Gogus2000,Gavriil2004}, fast radio bursts \citep{Gourdji2019}, network traffic \citep{Paxson1994,Singhai2007}, mining equipment failure \citep{LaRoche-Carrier2019}, earthquake aftershocks \citep{Peng2009}, and test response times \citep{VanDerLinden2006}.

\newpage
\section{Example fits for individual active regions} \label{app:perar}
\begin{figure}
    \centering
    \includegraphics[width=0.8\textwidth]{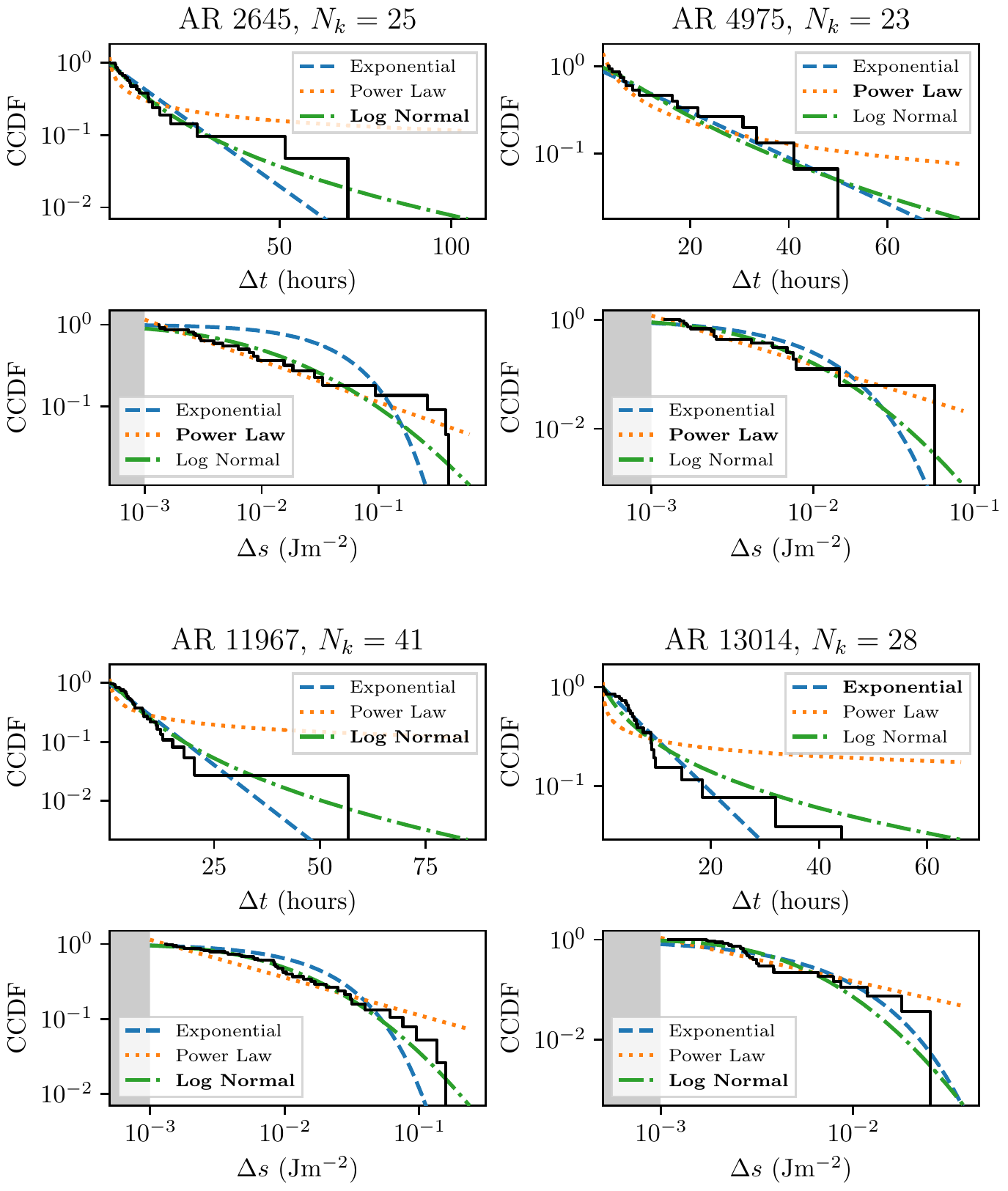}
    \caption{Complementary cumulative distribution functions (CCDFs) for four arbitrary but representative active regions with $N_k \geq 20$. The distribution that best describes the data in each panel (black stepped curve), according to the AICc, is in bold type-face in the legend for each panel. The grey regions in the bottom panels indicate $\Delta s \leq 10^{-3}\,$Jm$^{-2}$, i.e.~where the masking described in Section~\ref{sec:obscuration} is applied.}
    \label{fig:perar}
\end{figure}

The reader may be curious to examine some examples of the variety of distributions fitted to individual active regions. To this end, in Figure~\ref{fig:perar} we show the complementary cumulative distribution function (CCDF) for the flare waiting times and sizes from four arbitrary but representative active regions with $N_k \geq 20$. We display the distributions as CCDFs instead of PDFs to avoid binning the data. The black stepped curves indicate the empirical CCDF constructed from the \emph{GOES} catalog, while the colored curves are the various fits to the data, with parameters fixed to their maximum likelihood values. The distribution with bold type-face in the legend is the shape that best fits the data according to the AICc.

\end{document}